\newcommand{\CLASSINPUTbaselinestretch}{1.5}
\newcommand{\CLASSINPUTinnersidemargin}{1.01in}
\newcommand{\CLASSINPUTtoptextmargin}{1.65in}

\documentclass[12pt,draftclsnofoot,onecolumn,journal]{IEEEtran}

\usepackage{cite}%
\usepackage[T1]{fontenc}%
\usepackage[pdftex]{graphicx}%
\DeclareGraphicsExtensions{.pdf,.png}
\usepackage[cmex10]{amsmath}%
\usepackage{amssymb}%
\usepackage{amsthm}%
\usepackage[cmintegrals]{newtxmath}
\usepackage{bm}
\usepackage{algorithmicx}
\usepackage{array}%
\usepackage{mdwmath}
\usepackage{mdwtab}
\usepackage[caption=false,font=footnotesize]{subfig}%
\usepackage{url}
\usepackage{xcolor}

\providecommand{\U}[1]{\protect\rule{.1in}{.1in}}
\newtheorem{theorem}{Theorem}
\newtheorem*{theorem*}{Theorem}

\newtheorem{definition}{Definition}

\newtheorem{lemma}{Lemma}

\newcommand{\dif}{\mathop{}\!\mathrm{d}}

\DeclareMathOperator{\trace}{Tr}

\DeclareMathOperator{\diag}{diag}

\DeclareMathOperator{\supp}{supp}

\newcommand{\ket}[1]{\left| #1 \right>} 
\newcommand{\bra}[1]{\left< #1 \right|} 
\mathchardef\mhyphen="2D 

\IEEEoverridecommandlockouts

\begin{document}
\title{Covert Capacity of Bosonic Channels}

\author{Christos~N.~Gagatsos, Michael~S.~Bullock,~and Boulat~A.~Bash
\thanks{C.~N.~Gagatsos is with the College of Optical Sciences, University of Arizona, Tucson, AZ. M.~S.~Bullock is with the Electrical and Computer Engineering Department, University of Arizona, Tucson, AZ. B.~A.~Bash is with the Electrical and Computer Engineering Department, and the College of Optical Sciences, University of Arizona, Tucson, AZ.}
\thanks{CNG acknowledges the Office of Naval Research (ONR) MURI program on Optical Computing under grant no. N00014-14-1-0505. MSB and BAB were sponsored by the Army Research Office under Grant Number W911NF-19-1-0412.
This material is based upon work supported in part by the National Science Foundation under Grant No.~CCF-2006679.
The authors also acknowledge General Dynamics Mission Systems for supporting this research.}}

\maketitle

\begin{abstract}
We investigate the quantum-secure covert-communication capabilities of lossy thermal-noise bosonic channels, the quantum-mechanical model for many practical channels. 
We determine the expressions for the covert capacity of these channels: $L_{\text{no-EA}}$, when Alice and Bob share only a classical secret, and $L_{\text{EA}}$, when they benefit from entanglement assistance.
We find that entanglement assistance alters the fundamental scaling law for covert communication. 
Instead of $L_{\text{no-EA}}\sqrt{n}-r_{\text{no-EA}}(n)$, $r_{\text{no-EA}}(n)=o(\sqrt{n})$, entanglement assistance allows $L_{\text{EA}}\sqrt{n}\log n-r_{\text{EA}}(n)$, $r_{\text{EA}}(n)=o(\sqrt{n}\log n)$, covert bits to be transmitted reliably over $n$ channel uses.

\end{abstract}

\section{Introduction}
In contrast to standard information security methods (e.g., encryption, information-theoretic secrecy, and quantum key distribution (QKD)) that protect the transmission's content from unauthorized access, \emph{covert} or \emph{low probability of detection/intercept (LPD/LPI)} communication \cite{bash12sqrtlawisit,bash13squarerootjsacnonote,bash15covertcommmag} prevents  adversarial detection of transmissions in the first place.
The covertness requirement constrains the transmission power averaged over the blocklength $n$ to  $\propto 1/\sqrt{n}$, where the power is either measured directly in watts \cite{bash12sqrtlawisit,bash13squarerootjsacnonote} and mean photon number \cite{bash15covertbosoniccomm,bullock20discretemod} output by a physical transmitter, or indirectly, as the frequency of non-zero symbol transmission over the discrete classical \cite{bloch15covert,wang15covert} and quantum \cite{azadeh16quantumcovert-isitarxiv,wang16cq-srlconverse} channels.

For many channels, including classical additive white Gaussian noise (AWGN) \cite{bash12sqrtlawisit,bash13squarerootjsacnonote}, and discrete memoryless channels (DMCs) \cite{bloch15covert,wang15covert}, the power constraint prescribed by the covertness requirement imposes the \emph{square root law} (SRL): no more than $L\sqrt{n}$ covert bits can be transmitted reliably in $n$ channel uses. 
We call constant $L$ the \emph{covert capacity} of a channel, since it only depends on the channel parameters and captures a fundamental limit.
Attempting to transmit more results in either detection by the adversary with high 
  probability as $n\rightarrow\infty$, or unreliable transmission.
Even though the Shannon capacity \cite{cover02IT} of such channels is zero (since 
  $\lim_{n\rightarrow\infty}\frac{L\sqrt{n}}{n}=0$), 
  the SRL allows reliable transmission of a significant number of covert bits for large $n$.

\begin{figure}
	\centering
	\includegraphics[width=0.28\textwidth]{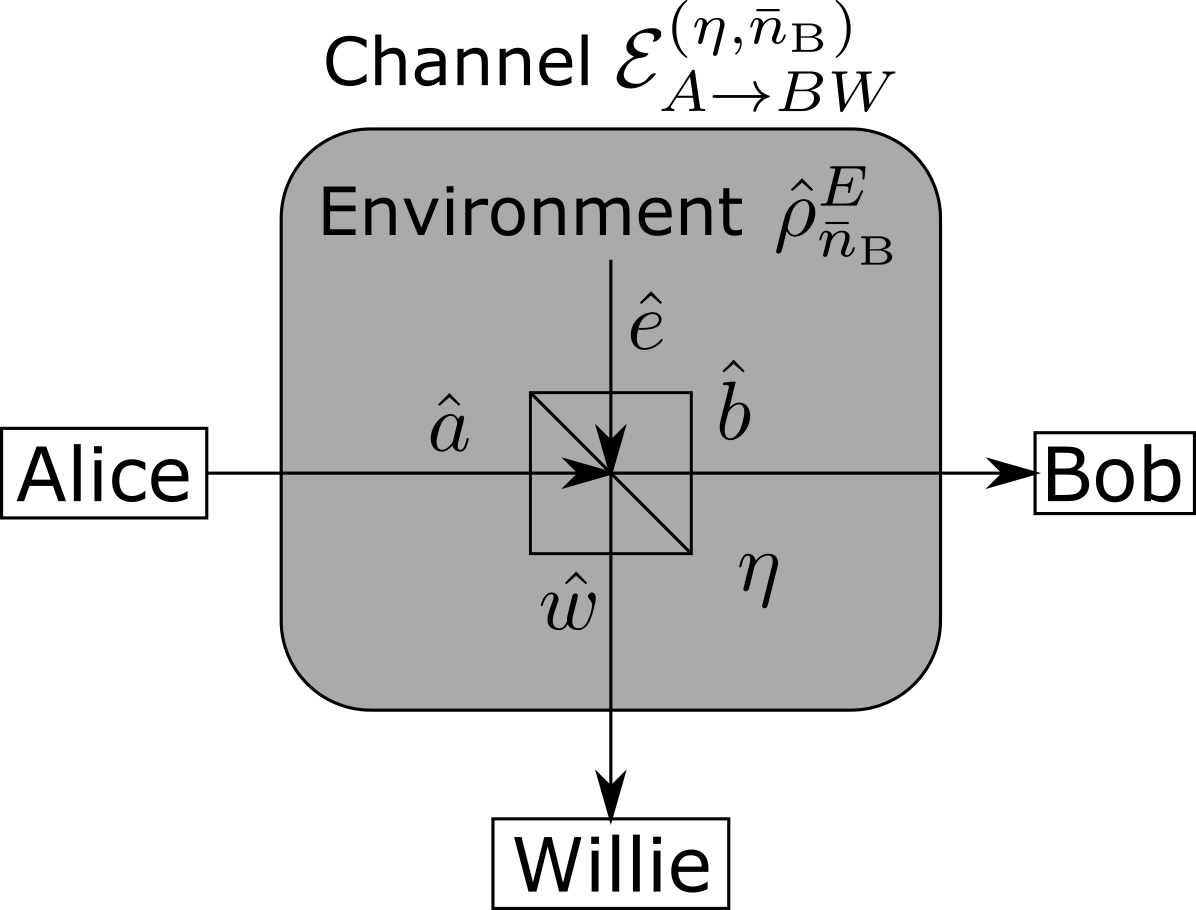}
	\caption{Single-mode bosonic channel $\mathcal{E}^{(\eta,\bar{n}_{\rm B})}_{A\to BW}$ 
	modeled by a beamsplitter with transmissivity $\eta$ and an environment 
	injecting a thermal state $\hat{\rho}_{\bar{n}_{\rm B}}$ with mean photon
	number $\bar{n}_{\rm B}$. $\hat{a}$, $\hat{e}$, $\hat{b}$, and $\hat{w}$ 
	label input/output modal annihilation operators.}
	\label{fig:optchannel}
\end{figure}

To date, the focus has been on classical covert communication. 
However, quantum mechanics governs the fundamental laws of nature, and quantum information theory \cite{nielsen00quantum,wilde16quantumit2ed} is required to determine the ultimate limits of any communications system.
Here we focus on the lossy thermal noise bosonic channel depicted in Fig.~\ref{fig:optchannel}, called the bosonic channel for brevity, and formally described in Section \ref{sec:channel_model}.
The bosonic channel is a quantum-mechanical model of many practical channels (including optical, microwave, and radio frequency (RF)).
This channel is parametrized by the power coupling (transmissivity) $\eta$ between the transmitter Alice and the intended receiver Bob, and the mean photon number $\bar{n}_{\rm B}$ per mode injected by the thermal environment, where a single spatial-temporal-polarization mode is our fundamental transmission unit.
We call a covert communication system \emph{quantum secure} when it is robust against an adversary Willie who not only knows the transmission parameters (including the start time, center frequency, duration, and bandwidth), but also has access to all the transmitted photons that are not captured by Bob, as well as arbitrary quantum information processing resources (e.g., joint detection measurement, quantum memory, and quantum computing).
While our approach is motivated by the security standards from the QKD literature, covertness demands a different set of assumptions.
We require excess noise that is not under Willie's control (e.g., the unavoidable thermal noise from the blackbody radiation at the center wavelength of transmission and the receiver operating temperature).
This assumption is not only well-grounded in practice, but also necessary for covertness, as the transmissions cannot be hidden from quantum-capable Willie that fully controls noise on the channel \cite[Th.~1]{bash15covertbosoniccomm},\cite{tahmasbi19covertqkd}.
Finally, we assume that Alice and Bob share a resource that is inaccessible by Willie. 
This enables covertness irrespective of channel conditions, as well as substantially increases the number of reliably-transmissible covert bits when the resource is an entangled quantum state.

In \cite{bullock20discretemod} we develop an expression for the maximum mean photon number $\bar{n}_{\rm S}$ that Alice can transmit under the aforementioned quantum-secure covertness conditions.
We also present the expression for the covert capacity $L$ for the bosonic channel and argue that it is achievable using a random coding scheme.
However, \cite{bullock20discretemod} focuses on the prescription for maintaining covertness of a transmission, with the capacity proofs left out.
Here, we fill in this gap by rigorously examining the coding limits for covert communication over the bosonic channels.

\begin{figure}
	\centering
	\includegraphics[width=0.95\textwidth]{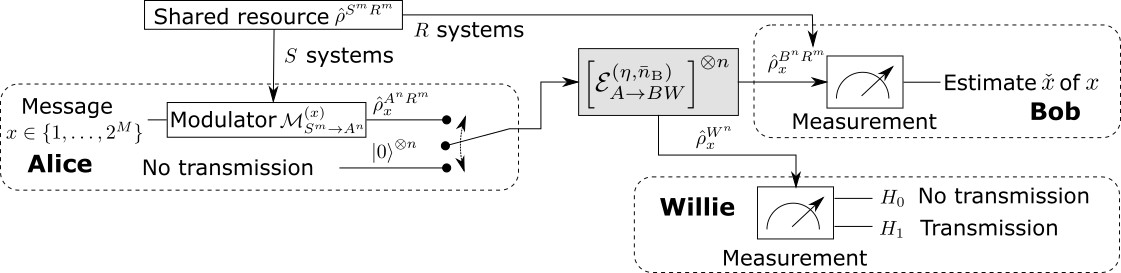}
	\caption{Covert communication over the bosonic channel.
	Alice has a bosonic channel, depicted in 
  Fig.~\ref{fig:optchannel}, to receiver Bob and adversary Willie. 
  Alice and Bob share a bipartite resource state $\hat{\rho}^{S^mR^m}$ that is inaccessible by Willie and may or may not be entangled.
  Alice uses her share of $\hat{\rho}^{S^mR^m}$ in $S$ systems to encode message $x$ with blocklength $n$ code, and chooses whether to
  transmit it using $\mathcal{E}^{(\eta,\bar{n}_{\rm B})}_{A\to BW}$ $n$ times.
  Willie observes his channel from Alice to determine whether she is quiet 
  (null hypothesis $H_0$) or not (alternate hypothesis $H_1$). 
  A covert communication system must ensure that any detector Willie uses is close to 
  ineffective (i.e., a random guess between the hypotheses), while allowing Bob to 
  reliably decode the message (if one is transmitted).}
	\label{fig:setup}
\end{figure}

Our main contribution is the analysis of the covert communication system depicted in Fig.~\ref{fig:setup} and formally described in Sec.~\ref{sec:system}, with and without an entangled resource state shared by Alice and Bob.  
Since entanglement assistance gain manifests only when $\bar{n}_{\rm S}\to0$ and $\bar{n}_{\rm B}>0$, we expect it to benefit covert communication. 
We find that entanglement assistance, in fact, alters the fundamental scaling law of covert communication.
Using the asymptotic notation defined in Sec.~\ref{sec:asym}:
\begin{enumerate}
\item We show that without entanglement assistance, the SRL has a standard form: $M_{\text{no-EA}}=L_{\text{no-EA}}\sqrt{n}-r_{\text{no-EA}}(n)$, $r_{\text{no-EA}}(n)=o(\sqrt{n})$, covert bits transmissible reliably over $n$ channel uses. 
Our second-order bound is similar to classical \cite{tahmasbi19covertdmc2ndorder}: $M_{\text{no-EA}}\geq L_{\text{no-EA}}\delta\sqrt{n}+K_{\text{no-EA}}\Phi^{-1}(\epsilon)n^{1/4}+\mathcal{O}(n^{n/8})$, where $\epsilon$ is the average decoding error probability and $\Phi^{-1}(x)$ is the inverse-Gaussian cumulative distribution function.\footnote{Note that $\Phi^{-1}(x)<0$ for $0<x<\frac{1}{2}$.}
We also show that quadrature phase shift keying (QPSK) modulation achieves the same constants $L_{\text{no-EA}}$ and $K_{\text{no-EA}}$ as the optimal Gaussian modulation.
\item We show that with entanglement assistance, the scaling law becomes $M_{\text{EA}}=L_{\text{EA}}\sqrt{n}\log n-r_{\text{EA}}(n)$, $r_{\text{EA}}(n)=o(\sqrt{n}\log n)$.
We derive the expression for the optimal constant $L_{\text{EA}}$ and the second-order bound.\footnote{Our fundamental information unit is a bit and $\log x$ indicates the binary logarithm, while $\ln x$ is the natural logarithm.} 
While a practical-receiver structure that achieves $L_{\text{EA}}$ is an open problem, we discuss a design \cite{guha20eajdr-isit} that achieves the $\mathcal{O}(\sqrt{n}\log n)$ scaling law, albeit with a constant $\approx \frac{L_{\text{EA}}}{2}$.
\end{enumerate}

Next, we present the mathematical prerequisites, including the asymptotic notation, the channel and system models, the formal definitions of covertness and reliability, and the bounds we need.  We state and prove our results in Sec.~\ref{sec:results}. We conclude with the discussion of future work in Sec.~\ref{sec:discussion}, including investigating the shared resource state size, the entanglement-assisted receiver design for covert communication, and the possible relationship of the scaling law for entanglement-assisted covert communication to a corner case in classical and non-entanglement assisted classical-quantum covert communication.

\section{Prerequisites}
\label{sec:prerequisites}

\subsection{Asymptotic notation}
\label{sec:asym}
We use the standard asymptotic notation \cite[Ch.~3.1]{clrs2e}, where 
 $f(n)=\mathcal{O}(g(n))$ denotes an asymptotic
  upper bound on $f(n)$ (i.e.~there exist constants $m,n_0>0$ such 
  that $0\leq f(n)\leq m g(n)$ for all $n\geq n_0$) and
  $f(n)=o(g(n))$ denotes an upper bound on $f(n)$ that is not 
  asymptotically tight (i.e.~for any constant $m>0$, there exists constant 
  $n_0>0$ such that $0\leq f(n)<m g(n)$ for all $n\geq n_0$).
We note that $f(n)=\mathcal{O}(g(n))$ is equivalent to $\limsup_{n\to\infty}\left|\frac{f(n)}{g(n)}\right|<\infty$ and $f(n)=o(g(n))$ is equivalent to $\lim_{n\to\infty}\frac{f(n)}{g(n)}=0$.

\subsection{Channel model}
\label{sec:channel_model}
We focus on a single-mode lossy thermal noise bosonic channel ${\cal E}^{(\eta,\bar{n}_{\rm B})}_{A\to BW}$ in Fig.~\ref{fig:optchannel}.
It quantum-mechanically describes the transmission of a single 
  (spatio-temporal-polarization) mode of the electromagnetic field at a given
  transmission wavelength (such as optical or microwave) over linear loss and
  additive Gaussian noise (such as noise stemming from blackbody radiation).
Here, we introduce the bosonic channel briefly, deferring the details to \cite{agarwal12quantumoptics,scully97quantumoptics,
 orszag16quantumotpics,shapiro16qoc}.
The attenuation in the Alice-to-Bob channel is modeled by a beamsplitter with transmissivity (fractional power coupling) $\eta$.
The input-output relationship between the bosonic modal annihilation operators of
  the beamsplitter, 
  ${\hat b} = \sqrt{\eta}{\hat a} + {\sqrt{1-\eta}}{\hat e}$,  
  requires the ``environment'' mode ${\hat e}$ to ensure that the commutator
  $\left[ {\hat b}, {\hat b}^\dagger \right]=1$, and to preserve the Heisenberg
  uncertainty law of quantum mechanics. 
On the contrary, classical power attenuation is described by 
  $b = {\sqrt{\eta}} a$, where $a$ and $b$ are complex amplitudes of input and
  output mode functions. 
Bob captures a fraction $\eta$ of Alice's transmitted photons, while Willie has access to the remaining $1-\eta$ fraction.
We model noise by mode $\hat e$ being in a zero-mean thermal state 
  $\hat{\rho}_{\bar{n}_{\rm B}}$, respectively expressed in the coherent state (quantum description of ideal laser light) and Fock
  (photon number) bases as follows:
\begin{align}
\label{eq:thermal}\hat{\rho}_{\bar{n}_{\rm B}}&=\frac{1}{\pi \bar{n}_{\rm B}}\int_{\mathbb{C}}\exp\left[-\frac{|\alpha|^2}{\bar{n}_{\rm B}}\right]\dif^2\alpha\ket{\alpha}\bra{\alpha}=\sum_{k=0}^\infty t_k(\bar{n}_{\rm B})\ket{k}\bra{k}, 
\end{align}
where
\begin{align}
\label{eq:tn}t_k(\bar{n})&=\frac{\bar{n}^k}{(1+\bar{n})^{k+1}}
\end{align}  
and $\bar{n}_{\rm B}$ is the mean photon number per mode injected by 
  the environment.

\subsection{System model}
\label{sec:system}
The covert communication framework is depicted in Fig.~\ref{fig:setup}.
Our fundamental transmission unit is the field mode described above.
We assume a discrete-time model with $n=2TW$ modes available to Alice and Bob. 
$TW$ is the number of orthogonal temporal modes, which is the product of 
  the transmission duration $T$ (in seconds) and the optical
  bandwidth $W$ (in Hz) of the source around its center frequency, and
  the factor of two corresponds to the use of both orthogonal polarizations.
The orthogonality of the available modes results in the bosonic channel ${\cal E}^{(\eta,\bar{n}_{\rm B})}_{A\to BW}$ being memoryless.
Alice and Bob have access to a bipartite resource state $\hat{\rho}^{S^mR^m}$ occupying $m$ systems $S$ at Alice and $R$ at Bob.
Correlations between parts of $\hat{\rho}^{S^mR^m}$ in systems $S$ and $R$ can either be classical or quantum, resulting in either a classical-quantum or an entangled state $\hat{\rho}^{S^mR^m}$. 
The latter allows entanglement-assisted communication.

\subsection{Coding and reliability}
\label{sec:reliability}
Alice desires to transmit one of $2^M$ equally-likely $M$-bit messages $x\in\left\{1,\ldots,2^M\right\}$ covertly to Bob using $n$ available modes of the bosonic channel ${\cal E}^{(\eta,\bar{n}_{\rm B})}_{A\to BW}$ and her share of the resource state $\hat{\rho}^{S^mR^m}$.
Her encoder is a set of encoding channels $\left\{\mathcal{M}^{(x)}_{S^m\to A^n}\right\}$.
Alice encodes message $x$ by acting on $m$ systems $S$ of $\hat{\rho}^{S^mR^m}$ with $\mathcal{M}^{(x)}_{S^m\to A^n}$, transforming $\hat{\rho}^{S^mR^m}$ to $\hat{\rho}^{A^nR^m}_x=\mathcal{M}^{(x)}_{S^m\to A^n}(\hat{\rho}^{S^mR^m})$.
Transmission of the resulting $n$ systems $A$ over $n$ uses of ${\cal E}^{(\eta,\bar{n}_{\rm B})}_{A\to BW}$ results in Bob receiving the state $\hat{\rho}^{B^nR^m}_x=\trace_{W^n}\left[\left[{\cal E}^{(\eta,\bar{n}_{\rm B})}_{A\to BW}\right]^{\otimes n}\left(\mathcal{M}^{(x)}_{S^m\to A^n}(\hat{\rho}^{S^mR^m})\right)\right]$, where $\hat{\rho}^A=\trace_B\left[\hat{\rho}^{AB}\right]$ denotes the partial trace over system $B$.
Bob decodes $x$ by applying a positive operator-valued measure (POVM) $\left\{\Lambda^{(x)}_{B^nR^m}\right\}$ to $\hat{\rho}^{B^nR^m}_x$.
Denoting by $X$ and $\check{X}$ the respective random variables corresponding to Alice's message and Bob's estimate of it, the average decoding error probability is:
\begin{align}
P_{\rm e}=\frac{1}{2^M}\sum_{x=1}^{2^M}P(\check{X}\neq x|X=x),
\end{align}
where $P(\check{X}\neq x|X=x)=\trace\left[\left(\hat{I}-\Lambda^{(x)}_{B^nR^m}\right)\hat{\rho}^{B^nR^m}_x\right]$.
We call the communication system \emph{reliable} if, for any $\epsilon\in(0,1)$, there exists $n$ large enough with a corresponding resource state $\hat{\rho}^{S^mR^m}$, encoder $\left\{\mathcal{M}^{(x)}_{S^m\to A^n}\right\}$, and decoder POVM $\left\{\Lambda^{(x)}_{B^nR^m}\right\}$, such that $P_{\rm e}\leq\epsilon$.

\subsection{Quantum-secure covertness}
\label{sec:hyptest}
As is standard in information theory of covert communication, we assume that Willie cannot access $\hat{\rho}^{S^mR^m}$, although he knows how it is generated.
To be \emph{quantum secure}, a covert communication system has to prevent the detection of Alice's transmission by Willie, who has access to all transmitted photons that are not received by Bob and arbitrary quantum resources.
Thus, the quantum state $\hat{\rho}^{W^n}_1=\sum_{x=1}^{2^M}\frac{1}{2^M}\trace_{B^nR^m}\left[\left[{\cal E}^{(\eta,\bar{n}_{\rm B})}_{A\to BW}\right]^{\otimes n}\left(\mathcal{M}^{(x)}_{S^m\to A^n}(\hat{\rho}^{S^mR^m})\right)\right]$, observed by Willie when Alice is transmitting, has to be sufficiently similar to the product thermal state $\hat{\rho}_{\eta\bar{n}_{\rm B}}^{\otimes n}$ that describes the noise observed when she is not.
We call a system \emph{covert} if, for any $\delta>0$ and $n$ large enough, 
  $D\left(\hat{\rho}^{W^n}_1\|\hat{\rho}_{\eta\bar{n}_{\rm B}}^{\otimes n}\right)\leq\frac{\delta}{\log e}$.
Arbitrarily small $\delta>0$ implies that the performance of a quantum-optimal detection scheme is arbitrarily close to that of a random coin flip through quantum Pinsker's inequality \cite[Th.~10.8.1]{wilde16quantumit2ed}.  
The properties of both classical and quantum relative entropy are highly attractive for mathematical proofs, and were used to analyze covert communication \cite{bash12sqrtlawisit, bash13squarerootjsacnonote, bash15covertcommmag, bash15covertbosoniccomm, bullock20discretemod, bloch15covert, wang15covert, azadeh16quantumcovert-isitarxiv,wang16cq-srlconverse}.
We discuss the significance of the quantum relative entropy in \cite[Sec.~II.B]{bullock20discretemod}.
The maximum mean photon number per mode $\bar{n}_{\rm S}$ that Alice can transmit under the covertness constraint is \cite{bullock20discretemod}:
\begin{align}  
\label{eq:ns}\bar{n}_{\rm S}&=\frac{\sqrt{\delta} c_{\rm cov}}{\sqrt{n}},
\end{align}
where
\begin{align}
\label{eq:c_cov}c_{\rm cov}&=\frac{\sqrt{2\eta\bar{n}_{\rm B}(1+\eta\bar{n}_{\rm B})}}{1-\eta}.
\end{align}
When the exact values for the environment mean photon number per mode $\bar{n}_{\rm B}$ and the transmissivity $\eta$ are unknown, Planck's law and the diffraction-limited propagation model provide a useful lower bound.
Coherent-state modulation using the continuous-valued complex Gaussian distribution \cite[Th.~2]{bash15covertbosoniccomm} and practical QPSK scheme \cite[Th.~2]{bullock20discretemod} achieve the constant \eqref{eq:c_cov}.

While quantum resources, such as entanglement shared between Alice and Bob, or quantum states lacking a semiclassical description (e.g., squeezed light) do not improve signal covertness, the quantum methodology allows covertness without assumptions of adversary's limits, other than the laws of physics.
However, the square root scaling in \eqref{eq:ns} holds even when Willie uses readily-available devices such as noisy photon counters \cite[Th.~5]{bash15covertbosoniccomm}, with a constant larger than $c_{\rm cov}$.
Nevertheless, here we show that quantum resources---specifically, entanglement assistance---allow the transmission of significantly more covert bits.
Next, we discuss the finite blocklength capacity bounds that we use in our proofs.

\subsection{Finite blocklength capacity bounds for bosonic channels}
One can obtain the converse results for covert communication using the standard channel coding theorems.
However, covertness introduces the dependence of the mean photon number per mode $\bar{n}_{\rm S}$ on the blocklength $n$ in \eqref{eq:ns}.
This complicates both classical and quantum achievability proofs by rendering invalid the conditions for employing standard results such as the asymptotic equipartition property.
Classical results \cite{bloch15covert,wang15covert} overcome this issue using the information spectrum methods \cite{han93resolvability,han03infspectrum,polyanskiy10finiteblocklength}.
However, until recently, quantum information spectrum approaches \cite{hayashi2003general,nagaoka2007information} have been limited to channels with output quantum states living in the finite-dimensional Hilbert space, which is not the case for bosonic channels.
We now rehash a lower bound on the second-order coding rate from \cite{oskouei18unionbound,wilde17ppc} that is based on the new quantum union bound \cite{oskouei18unionbound}.

Define quantum relative entropy $D(\hat{\rho}\|\hat{\sigma})$ between states $\hat{\rho}$ and $\hat{\sigma}$, and its second, third, and fourth absolute central moments as follows:
\begin{align}
\label{eq:qre}D(\hat{\rho}\|\hat{\sigma})&=\trace\left[\hat{\rho}\log\hat{\rho}-\hat{\rho}\log\hat{\sigma}\right]\\
\label{eq:qrev}V(\hat{\rho}\|\hat{\sigma})&=\trace\left[\hat{\rho}\left|\log\hat{\rho}-\log\hat{\sigma}-D(\hat{\rho}\|\hat{\sigma})\right|^2\right]\\
\label{eq:qT}T(\hat{\rho}\|\hat{\sigma})&=\trace\left[\hat{\rho}\left|\log\hat{\rho}-\log\hat{\sigma}-D(\hat{\rho}\|\hat{\sigma})\right|^3\right]\\
\label{eq:qQ}Q(\hat{\rho}\|\hat{\sigma})&=\trace\left[\hat{\rho}\left|\log\hat{\rho}-\log\hat{\sigma}-D(\hat{\rho}\|\hat{\sigma})\right|^4\right],
\end{align}
where $V(\hat{\rho}\|\hat{\sigma})$ is quantum relative entropy variance.
The finite blocklength capacity of a memoryless classical-quantum channel described in Sec.~\ref{sec:system} is characterized as follows:
\begin{lemma}\label{lemma:Mbound}
Suppose that the channel from Alice to Bob is memoryless, such that over $n$ uses $\mathcal{N}_{A^n\to B^n}=\left(\mathcal{N}_{A\to B}\right)^{\otimes n}$.
There exists a coding scheme that employs a shared resource state $\rho^{S^mR^m}$ to transmit $M$ bits over $n$ uses of $\mathcal{N}_{A\to B}$ with arbitrary decoding error probability $\epsilon$ for a sufficiently large $n$ and $m$, such that:
\begin{align}
\label{eq:Mbound}M&\geq n D\left(\hat{\rho}^{BR}\middle\|\hat{\rho}^{B}\otimes\hat{\rho}^{R}\right)+\sqrt{nV\left(\hat{\rho}^{BR}\middle\|\hat{\rho}^{B}\otimes\hat{\rho}^{R}\right)}\Phi^{-1}(\epsilon)-C_n,
\end{align}
where $C_n=\frac{C_{\text{B-E}}}{\sqrt{2\pi}}\frac{\left[Q\left(\hat{\rho}^{BR}\middle\|\hat{\rho}^{B}\otimes\hat{\rho}^{R}\right)\right]^{3/4}}{V\left(\hat{\rho}^{BR}\middle\|\hat{\rho}^{B}\otimes\hat{\rho}^{R}\right)}+\frac{\sqrt{V\left(\hat{\rho}^{BR}\middle\|\hat{\rho}^{B}\otimes\hat{\rho}^{R}\right)}}{\sqrt{2\pi}}+\log (4\epsilon n)$, $C_{\text{B-E}}$ is the Berry-Esseen constant satisfying $0.40973\leq C_{\text{B-E}}\leq 0.4784$, $\hat{\rho}^{BR}$ is Bob's marginal state for the output of a single channel use, and $\Phi^{-1}(x)$ is the inverse-Gaussian distribution function. 
\end{lemma}
\begin{IEEEproof}[Proof \cite{oskouei18unionbound,wilde17ppc}]
Suppose Alice and Bob have access to $m=n2^M$ bipartite systems $SR$ containing $m$ copies of the resource state $\hat{\rho}^{SR}$, with Alice restricted to system $S$ and Bob to system $R$. 
Denote by $\hat{\rho}^{S_kR_k}$ the state $\hat{\rho}^{SR}$ in the $k^{\text{th}}$ bipartite system $S_kR_k$, $k=1,\ldots,m$. 
Alice and Bob agree to divide these $m$ systems into $2^M$ non-overlapping $n$-system subsets, each mapping to a message $x\in\left\{1,\ldots,2^M\right\}$.  Denote the corresponding subsets of system indexes by $\mathcal{X}(x)$.
The encoding channel is thus $\mathcal{M}^{(x)}_{S^m\to A^n}\left(\left(\hat{\rho}^{SR}\right)^{\otimes m}\right)=\trace_{S_k,k\notin\mathcal{X}(x)}\left[\left(\hat{\rho}^{SR}\right)^{\otimes m}\right]$.
Alice sends $x$ to Bob by transmitting the corresponding codeword over $n$ uses of $\mathcal{N}_{A\to B}$.
The authors of \cite{oskouei18unionbound,wilde17ppc} call this scheme \emph{position-based coding}.
Bob's received state is $\hat{\rho}^{B^nR^m}=\bigotimes_{k\in\mathcal{X}(x)}\hat{\rho}^{B_kR_k}\bigotimes_{k\notin\mathcal{X}(x)}\trace_{S_k}\left[\hat{\rho}^{S_kR_k}\right]$, where $\hat{\rho}^{B_kR_k}=\mathcal{N}_{A\to B}\left(\hat{\rho}^{S_kR_k}\right)$.
Bob constructs $2^M$ binary projective measurements corresponding to each message, and applies them sequentially to $\hat{\rho}^{B^nR^m}$.
This operation, resembling a matched filter from classical communication, is called \emph{sequential decoding} in \cite{oskouei18unionbound,wilde17ppc}.
Its analysis in \cite[Sec.~5]{oskouei18unionbound} proves that 
\begin{align}
\label{eq:hyptestrelent}M&\geq D_{\rm H}^{\epsilon-\zeta}\left(\left(\hat{\rho}^{BR}\right)^{\otimes n}\middle\|\left(\hat{\rho}^{B}\right)^{\otimes n}\otimes\left(\hat{\rho}^{R}\right)^{\otimes n}\right)-\log\left(4\epsilon/\zeta^2\right),
\end{align}
where $D_{\rm H}^{\epsilon}\left(\hat{\rho}\middle\|\hat{\sigma}\right)$ is the hypothesis testing relative entropy \cite{buscemi10quantcap,wang12oneshot} defined in \cite[Eq.~(5.2)]{oskouei18unionbound} and $\zeta\in(0,\epsilon)$.
Specifically, \cite[Corr.~8]{oskouei18unionbound} yields \eqref{eq:hyptestrelent} when $\hat{\rho}^{SR}$ is classical-quantum  and \cite[Th.~6]{oskouei18unionbound} when it is entangled.
The standard steps for deriving the second-order rate bounds in the proof of \cite[Prop.~13]{oskouei18unionbound} up to \cite[Eq.~(A.24)]{oskouei18unionbound} yield:
\begin{align}
\label{eq:DHbound}D_{\rm H}^{\epsilon-\zeta}\left(\left(\hat{\rho}^{BR}\right)^{\otimes n}\middle\|\left(\hat{\rho}^{B}\right)^{\otimes n}\otimes\left(\hat{\rho}^{R}\right)^{\otimes n}\right)&\geq n D\left(\hat{\rho}^{BR}\middle\|\hat{\rho}^{B}\otimes\hat{\rho}^{R}\right)+\sqrt{nV\left(\hat{\rho}^{BR}\middle\|\hat{\rho}^{B}\otimes\hat{\rho}^{R}\right)}\Phi^{-1}\left(\epsilon-C_n^\prime\right),
\end{align}
where $C_n^{\prime}=\zeta+\frac{C_{\text{B-E}}T\left(\hat{\rho}^{BR}\middle\|\hat{\rho}^{B}\otimes\hat{\rho}^{R}\right)}{\sqrt{n\left[V\left(\hat{\rho}^{BR}\middle\|\hat{\rho}^{B}\otimes\hat{\rho}^{R}\right)\right]^3}}$.
Substituting \eqref{eq:DHbound} in \eqref{eq:hyptestrelent}, setting $\zeta=1/\sqrt{n}$, expanding $\Phi^{-1}\left(\epsilon-C_n^\prime\right)$ at $\epsilon$ using Lagrange's mean value theorem \cite[Th.~2.3]{lang97ugradanalysis}, upperbounding $e^{-x^2}\leq 1$, and employing the convexity argument in \cite[Eq.~(D.3)]{kaur17ubsecretkey} completes the proof.
\end{IEEEproof} 
In contrast to \cite{oskouei18unionbound}, Lemma \ref{lemma:Mbound} does not absorb the remainder terms of $C_n$ in asymptotic notation, making \eqref{eq:Mbound} exact.
This is done to account for the dependence of $\bar{n}_{\rm S}$ on $n$ imposed by the covertness constraint \eqref{eq:ns}.
Finally, note that skipping the convexity argument concluding the proof yields a tighter but analytically inconvenient bound using \eqref{eq:qT} instead of \eqref{eq:qQ}.

\section{Results}
\label{sec:results}
\subsection{Covert channel capacity}
\label{sec:covcap}
In classical and quantum information theory \cite{cover02IT,nielsen00quantum,wilde16quantumit2ed}, the channel capacity is measured in bits per channel use and is expressed as $C=\liminf_{n\to\infty}\frac{M}{n}$, 
where $M$ is the total number of reliably-transmissible bits in $n$ channel uses.
On the other hand, the power constraint \eqref{eq:ns} imposed by covert communication implies that $M=o(n)$ and that the capacity of the covert channel is zero.
Inspired by \cite{wang15covert}, we regularize the number of covert bits that are transmitted reliably without entanglement assistance by $\sqrt{n}$ and with entanglement assistance $\sqrt{n}\log n$, instead of $n$.
This approach allows us to state Definitions \ref{def:cap-no-EA} and \ref{def:cap-EA} of covert channel capacity and derive the results that follow.
For consistency with \cite{wang15covert}, we also normalize the capacity by the covertness parameter $\delta$, which we discuss in Section \ref{sec:hyptest}.

\subsection{Covert communication without entanglement assistance}
We define the capacity of covert communication over the bosonic channel when Alice and Bob do not have access to a shared entanglement source as follows:
\begin{definition}\label{def:cap-no-EA} The capacity of covert communication without entanglement assistance is:
\begin{align}
L_{\rm{no\mhyphen EA}}&\triangleq \liminf_{n\to\infty}\frac{M_{\rm{no\mhyphen EA}}}{\sqrt{\delta n}},
\end{align}
where $M_{\rm{no\mhyphen EA}}$ is the number of covert bits that are reliably transmissible in $n$ channel uses (modes), and $\delta$ parametrizes the desired covertness.
\end{definition}
The following theorem provides the expression for $L_{\text{no-EA}}$:
\begin{theorem}
\label{th:no-ea}
The covert capacity of the bosonic channel without entanglement assistance is
$L_{\rm{no\mhyphen EA}}=c_{\rm cov}c_{\rm{rel,no\mhyphen EA}}$, 
where $c_{\rm cov}$ is defined in \eqref{eq:c_cov} and $c_{\rm{rel,no\mhyphen EA}}=\eta\log\left(1+\frac{1}{(1-\eta)\bar{n}_{\rm B}}\right)$.
\end{theorem}

In order to prove Theorem \ref{th:no-ea}, we prove the following lemma:

\begin{lemma}
\label{lemma:noEAfiniteblocklengthn}
There exists a sequence of codes with covertness parameter $\delta$, blocklength $n$, size $2^M$, and average error probability $\epsilon$ that satisfies:
\begin{align}
\label{eq:MnoEA}M_{\rm{no\mhyphen EA}}&\geq L_{\rm{no\mhyphen EA}}\sqrt{\delta n}+K_{\rm{no\mhyphen EA}}\Phi^{-1}(\epsilon)n^{1/4}+\mathcal{O}\left(n^{1/8}\right), 
\end{align}
where $K_{\rm{no\mhyphen EA}}=\sqrt{c_{\rm{cov}}\sqrt{\delta}(1+2(1-\eta)\bar{n}_{\rm B})}c_{\rm{rel,no\mhyphen EA}}$.
\end{lemma}

\begin{IEEEproof}
Alice and Bob follow the construction in the proof of Lemma \ref{lemma:Mbound} and generate a random codebook $\mathcal{C}=\{\mathbf{c}(x),x=1,\ldots,2^M\}$ mapping $M$-bit input blocks to $n$-symbol codewords.
Each $\mathbf{c}(x)\in\mathcal{Q}^n$ is generated according to $p(\mathbf{c})=\prod_{k=1}^np_{\rm q}(c_k)$, where $\mathcal{Q}=\{a,ja,-a,-ja\}$ is the QPSK alphabet and $p_{\rm q}(y)=\frac{1}{4}$ is the uniform distribution over it.
We set $a=\sqrt{\bar{n}_{\rm S}}$, where $\bar{n}_{\rm S}$ is defined in \eqref{eq:ns}.
We assume that $\mathcal{C}$ is shared between Alice and Bob before transmission, and is kept secret from Willie.
Product coherent states are modulated with amplitudes corresponding to the symbols in each codeword $\mathbf{c}(x)$: $\hat{\rho}^{A^n}_x=\bigotimes_{k=1}^n\ket{c_k(x)}\bra{c_k(x)}^A$.
Thus, Alice transmits the maximum mean photon number that maintains covertness \cite[Th.~2]{bullock20discretemod}.

The shared resource state $\hat{\rho}^{S^mR^m}=\left(\hat{\rho}^{SR}\right)^{\otimes m}$ is the random codebook $\mathcal{C}$ modulated by coherent states.
Thus, $\hat{\rho}^{SR}=\sum_{y\in\mathcal{Q}}p_{\rm q}(y)\ket{y}\bra{y}^{A}\otimes\ket{y}\bra{y}^{R}$ is a classical-quantum state, system $A$ is in a coherent state $\ket{y}^A$, and system $R$ is in one of the orthonormal states $\ket{y}^{R}$ corresponding to QPSK symbol index.
Alice's position-based encoder then selects the systems $S$ corresponding to the $n$-symbol codeword for message $x$, and discards the rest.
Bob employs the sequential decoding strategy described in \cite[Sec.~5]{oskouei18unionbound} and \cite[Sec.~3]{wilde17ppc}.

Since the propagation of a coherent state $\ket{\alpha}$ through the bosonic channel ${\cal E}_{A\to B}^{(\eta,\bar{n}_{\rm B})}$ induces a displaced thermal state $\hat{\rho}_{(1-\eta)\bar{n}_{\rm B}}(\eta\alpha)\equiv\hat{\rho}_{\rm T}(\alpha)$ in Bob's output port,
the received state is $\hat{\rho}^{BR}=\sum_{y\in\mathcal{Q}}p_{\rm q}(y)\hat{\rho}^B_{\rm T}(y)\otimes\ket{y}\bra{y}^{R}$, where displaced thermal states $\left\{\hat{\rho}^B_{\rm T}(y),y\in\mathcal{Q}\right\}$ form an ensemble corresponding to the transmission of QPSK symbols.
Letting $\hat{\bar{\rho}}^B\equiv\sum_{y\in\mathcal{Q}}p_{\rm q}(y)\hat{\rho}^B_{\rm T}(y)$, 
\begin{align}
\label{eq:chidef}D\left(\hat{\rho}^{BR}\middle\|\hat{\rho}^{B}\otimes\hat{\rho}^{R}\right)&=\chi\left(\left\{p_{\rm q}(y),\hat{\rho}^B_{\rm T}(y)\right\}\right)=S\left(\hat{\bar{\rho}}^B\right)-\sum_{y\in\mathcal{Q}}p_{\rm q}(y)S\left(\hat{\rho}^B_{\rm T}(y)\right)\\
V\left(\hat{\rho}^{BR}\middle\|\hat{\rho}^{B}\otimes\hat{\rho}^{R}\right)&=V_\chi\left(\left\{p_{\rm q}(y),\hat{\rho}^B_{\rm T}(y)\right\}\right)\nonumber\\
\label{eq:vchidef}&=\sum_{y\in\mathcal{Q}}p_{\rm q}(y)\left[V\left(\hat{\rho}^B_{\rm T}(y)\middle\|\hat{\bar{\rho}}^B\right)+\left[D\left(\hat{\rho}^B_{\rm T}(y)\middle\|\hat{\bar{\rho}}^B\right)\right]^2\right]-\left[\chi\left(\left\{p_{\rm q}(y),\hat{\rho}^B_{\rm T}(y)\right\}\right)\right]^2,
\end{align}
where the von Neumann entropy is 
\begin{align}
\label{eq:S}S(\hat{\rho})&=-\trace[\hat{\rho}\log\hat{\rho}],
\end{align}
while $\chi\left(\left\{p(x),\hat{\rho}_x\right\}\right)$ and $V_\chi\left(\left\{p(x),\hat{\rho}_x\right\}\right)$ are the Holevo information and its variance for ensemble $\left\{p(x),\hat{\rho}_x\right\}$.
The closed-form expressions for $\chi\left(\left\{p_{\rm q}(y),\hat{\rho}^B_{\rm T}(y)\right\}\right)$ and $V_\chi\left(\left\{p_{\rm q}(y),\hat{\rho}^B_{\rm T}(y)\right\}\right)$ are unknown, and we derive the Taylor series expansions at $\bar{n}_{\rm S}=0$ in Appendices \ref{app:chiQPSK} and \ref{app:vQPSK}:
\begin{align}
\label{eq:chiQPSK}\chi\left(\left\{p_{\rm q}(y),\hat{\rho}^B_{\rm T}(y)\right\}\right)&=\bar{n}_{\rm S}c_{\mathrm{rel},\text{no-EA}}+\mathcal{O}(\bar{n}_{\rm S}^2)\\
\label{eq:VchiQPSK}V_\chi\left(\left\{p_{\rm q}(y),\hat{\rho}^B_{\rm T}(y)\right\}\right)&=\bar{n}_{\rm S}(1+2(1-\eta)\bar{n}_{\rm B})c_{\mathrm{rel},\text{no-EA}}+\mathcal{O}(\bar{n}_{\rm S}^2).
\end{align}
In Appendix \ref{app:qQPSK} we show that $Q\left(\hat{\rho}^{BR}\middle\|\hat{\rho}^{B}\otimes\hat{\rho}^{R}\right)=\mathcal{O}\left(\bar{n}_{\rm S}\right)$.
We complete the proof by substituting $\bar{n}_{\rm S}$ from \eqref{eq:ns} and observing that $\frac{\left[Q\left(\hat{\rho}^{BR}\middle\|\hat{\rho}^{B}\otimes\hat{\rho}^{R}\right)\right]^{3/4}}{V\left(\hat{\rho}^{BR}\middle\|\hat{\rho}^{B}\otimes\hat{\rho}^{R}\right)}=\mathcal{O}\left(n^{1/8}\right)$ dominates the remainder $C_n$ in Lemma \ref{lemma:Mbound}.
\end{IEEEproof}

We are now ready to prove Theorem \ref{th:no-ea}.

\begin{IEEEproof}[Proof (Theorem \ref{th:no-ea})] \textbf{Achievability:}
Dividing both sides of \eqref{eq:MnoEA} by $\sqrt{n\delta}$ and taking the limit yields the achievable lower bound.

\textbf{Converse:} 
Let Alice and Bob have access to respective systems $S$ and $R$ of shared  infinite-dimensional bipartite classically-correlated resource state $\hat{\rho}^{S^mR^m}$, with $m$ arbitrary.
Consider a sequence of codes such that the decoding error probability $\epsilon_n\to0$ as $n\to\infty$.
Then:
\begin{align}
\label{eq:fano_noEA}M_{\text{no-EA}}(1-\epsilon_n)-1&\leq I\left(X^{(n)};\check{X}^{(n)}\right)\\
\label{eq:epni_bound}&\leq n C_{\chi}(\bar{n}_{\rm S};\eta,\bar{n}_{\rm B}),
\end{align}
where $I\left(X^{(n)};\check{X}^{(n)}\right)$ is the mutual information between random variables $X^{(n)}$ and $\check{X}^{(n)}$ corresponding to Alice's message and Bob's decoding of it, \eqref{eq:fano_noEA} follows from  Fano's inequality \cite[Th.~2.10.1]{cover02IT}, \eqref{eq:epni_bound} is the Holevo bound \cite[Th.~12.1]{nielsen00quantum}, \cite{holevo73bound} and $C_{\chi}(\bar{n}_{\rm S};\eta,\bar{n}_{\rm B})=g(\eta\bar{n}_{\rm S}+(1-\eta)\bar{n}_{\rm B})-g((1-\eta)\bar{n}_{\rm B})$ is the Holevo capacity of the bosonic channel from Alice to Bob $\mathcal{E}^{(\eta,\bar{n}_{\rm B})}_{A\to B}$ \cite{giovannetti2014ultimate} with
\begin{align}
\label{eq:g}g(x)&\equiv(1+x)\log(1+x)-x\log x.
\end{align}
The Taylor series expansion of $C_{\chi}(\bar{n}_{\rm S};\eta,\bar{n}_{\rm B})$ around $\bar{n}_{\rm S}=0$ in \eqref{eq:epni_bound} yields:
\begin{align}
C_{\chi}(\bar{n}_{\rm S};\eta,\bar{n}_{\rm B})&=\eta\bar{n}_{\rm S}\log\left(1+\frac{1}{(1-\eta)\bar{n}_{\rm B}}\right)-\frac{\eta^2\bar{n}_{\rm S}^2}{(2\ln 2)((1-\eta)\bar{n}_{\rm B}(1+(1-\eta)\bar{n}_{\rm B})}+o(\bar{n}_{\rm S}^3)\\
\label{eq:taylorbound}&\leq \eta\bar{n}_{\rm S}\log\left(1+\frac{1}{(1-\eta)\bar{n}_{\rm B}}\right),
\end{align}
where \eqref{eq:taylorbound} follows from Taylor's theorem with the remainder \cite[Ch.~V.3]{lang97ugradanalysis}.
Substituting \eqref{eq:taylorbound} and \eqref{eq:ns} in \eqref{eq:epni_bound} yields:
\begin{align}
\label{eq:MnoEAub}M_{\text{no-EA}}(1-\epsilon_n)-1&\leq\sqrt{n\delta} c_{\rm cov}\eta\log\left(1+\frac{1}{(1-\eta)\bar{n}_{\rm B}}\right).
\end{align}
Dividing both sides of \eqref{eq:MnoEAub} by $\sqrt{n\delta}$ and taking the limit yields the converse.
\end{IEEEproof}

\emph{Remark:} 
Since, to our knowledge, \cite[Corr.~8]{oskouei18unionbound} was proven only for the discrete inputs, Lemma \ref{lemma:Mbound} does not apply to the Gaussian ensemble of coherent states $\mathcal{G}=\left\{\frac{1}{\pi\bar{n}_{\rm S}}\exp\left[-\frac{|\alpha|^2}{\bar{n}_{\rm S}}\right],\ket{\alpha}\right\}$ directly.
However, since $\mathcal{G}$ achieves the Holevo capacity of the bosonic channel, we compare the information quantities for QPSK modulation in \eqref{eq:chiQPSK} and \eqref{eq:VchiQPSK} to the corresponding ones for $\mathcal{G}$.
Comparison of \eqref{eq:chiQPSK} and \eqref{eq:taylorbound} confirms the well-known fact \cite{lacerda17cohstateconstellations} that QPSK modulation achieves the Holevo capacity in the low signal-to-noise ratio (SNR) regime.
We calculate the Holevo information variance for $\mathcal{G}$ in Appendix \ref{app:QREVnonEA} and note that \eqref{eq:VchiQPSK} and \eqref{eq:AppVarGaussianApprox} have the same first term.
Thus, the QPSK modulation has the same finite blocklength performance as $\mathcal{G}$ in the low SNR regime.

\subsection{Entanglement-assisted covert communication}

Entanglement assistance increases the communication channel capacity  \cite{bsst02eacap,giovannetti03broadband}.
However, in most practical settings (including optical communication where noise level is low $\bar{n}_{\rm B}\ll1$ and microwave/RF communication where signal power is high $\bar{n}_{\rm S}\gg1$), the gain over the Holevo capacity without entanglement assistance is at most a factor of two.
The only scenario with a significant gain is when $\bar{n}_{\rm S}\to 0$ while $\bar{n}_{\rm B}>0$ \cite[App.~A]{guha20eajdr-isit}.
This is precisely the covert communication setting.
In fact, entanglement assistance alters the fundamental square root scaling law for covert communication, changing the normalization from $\sqrt{n}\log n$ to $\sqrt{n}$:

\begin{definition}\label{def:cap-EA} The capacity of covert communication with entanglement assistance is:
\begin{align}
L_{\rm{EA}}&\triangleq \liminf_{n\to\infty}\frac{M_{\rm{EA}}}{\sqrt{\delta n}\log{n}},
\end{align}
where $M_{\rm{EA}}$ is the number of covert bits that are reliably transmissible in $n$ channel uses (modes), and $\delta$ parametrizes the desired covertness.
\end{definition}
The following theorem provides the expression for $L_{\rm{EA}}$:
\begin{theorem}
\label{th:ea}
The covert capacity of the bosonic channel with entanglement assistance is 
$L_{\rm{EA}}=c_{\rm cov}c_{\rm{rel,EA}}$,
where $c_{\rm cov}$ is defined in \eqref{eq:c_cov} and $c_{\rm{rel,EA}}=\frac{\eta}{2(1+(1-\eta)\bar{n}_{\rm B})}$.
\end{theorem}
Thus, while quantum resources such as shared entanglement and joint detection receivers do not affect $\bar{n}_{\rm S}$, they dramatically impact the amount of information that can be covertly conveyed.
As in the proof of Theorem \ref{th:no-ea}, in order to prove Theorem \ref{th:ea}, we prove the following lemma:

\begin{lemma}
\label{lemma:EAfiniteblocklengthn}
There exists a sequence of codes with covertness parameter $\delta$, blocklength $n$, size $2^M$, and average error probability $\epsilon$ that satisfy:
\begin{align}
\label{eq:M_EAlb}M_{\rm{EA}}&\geq L_{\rm{EA}}\sqrt{\delta n}\log n+K_{\rm{EA}}\Phi^{-1}(\epsilon)n^{1/4}\log n+\mathcal{O}\left(n^{1/8}\log n\right),
\end{align}
where $K_{\rm{EA}}=\sqrt{c_{\rm{cov}}\sqrt{\delta} c_{\rm{rel},\rm{EA}}}$.
\end{lemma}

\begin{IEEEproof}[Proof (Lemma \ref{lemma:EAfiniteblocklengthn})]
Let the resource state be a tensor product $\left(\ket{\psi}^{SR}\right)^{\otimes m}$ of a two-mode squeezed vacuum (TMSV) states such that $m=n2^M$ and 
$\ket{\psi}^{SR}=\sum_{k=1}^\infty \sqrt{t_k(\bar{n}_{\rm S})}\ket{k}^S\ket{k}^{R}$, where $t_k(\bar{n})$ and $\bar{n}_{\rm S}$  are defined in \eqref{eq:tn} and \eqref{eq:ns}, respectively.
Alice and Bob use the position-based code \cite[Th.~6]{oskouei18unionbound} as in the proof of Lemma \ref{lemma:Mbound} and assign each message to $n$ TMSV states.
Alice transmits $n$ modes corresponding to message $x$ from her part of $\left(\ket{\psi}^{SR}\right)^{\otimes m}$, discarding the rest.
Willie has no access to Bob's system $R$.
Since $\trace_{R}\left[\ket{\psi}\bra{\psi}^{SR}\right]=\hat{\rho}_{\bar{n}_{\rm S}}$ is a thermal state, setting $\bar{n}_{\rm S}$ as in \eqref{eq:ns} ensures covertness \cite[Th.~2]{bash15covertbosoniccomm}.

Bob uses the sequential decoding from \cite[Th.~6]{oskouei18unionbound}, as in Lemma \ref{lemma:Mbound}.
In order to obtain the constants in \eqref{eq:M_EAlb}, first we note that $D\left(\hat{\rho}^{BR}\middle\|\hat{\rho}^{B}\otimes\hat{\rho}^{R}\right)=C_{\text{EA}}(\bar{n}_{\rm S};\eta,\bar{n}_{\rm B})$ \cite{giovannetti03broadband}, where
\begin{align}
\label{eq:ea_cap}C_{\text{EA}}(\bar{n}_{\rm S};\eta,\bar{n}_{\rm B})&=g(\bar{n}_{\rm S})+g(\eta\bar{n}_{\rm S}+(1-\eta)\bar{n}_{\rm B})-g(A_{+})-g(A_{-}),
\end{align}
 $A_{\pm}=\frac{B-1\pm (1-\eta)(\bar{n}_{\rm B}-\bar{n}_{\rm S})}{2}$, $B=\sqrt{(\bar{n}_{\rm S}+1+\eta\bar{n}_{\rm S}+(1-\eta)\bar{n}_{\rm B})^2-4\eta\bar{n}_{\rm S}(\bar{n}_{\rm S}+1)}$, and $g(x)$ is defined in \eqref{eq:g}.
The following expansion of $C_{\text{EA}}(\bar{n}_{\rm S};\eta,\bar{n}_{\rm B})$ around $\bar{n}_{\rm S}=0$:
\begin{align}
\nonumber C_{\text{EA}}(\bar{n}_{\rm S};\eta,\bar{n}_{\rm B})&=-\frac{\eta\bar{n}_{\rm S}\log \bar{n}_{\rm S}}{1+(1-\eta)\bar{n}_{\rm B}}-\bar{n}_{\rm S}\left(\frac{1}{\ln 2}+\frac{\eta}{1+(1-\eta)\bar{n}_{\rm B}}\log\left[1-\frac{1}{1+(1-\eta)\bar{n}_{\rm B}}\right]\right)\\
&\phantom{=}+\mathcal{O}(\bar{n}_{\rm S}^2\log \bar{n}_{\rm S})
\end{align}
yields $-\bar{n}_{\rm S}\log\bar{n}_{\rm S}$ as the dominant term in \eqref{eq:ea_cap}.

The expression for $L_{\rm EA}$ follows from the substitution of \eqref{eq:ns} in \eqref{eq:ea_cap} and the limit:
\begin{align}
\label{eq:L_EA_limit}\lim_{n\to\infty}\frac{nC_{\text{EA}}\left.(\bar{n}_{\rm S};\eta,\bar{n}_{\rm B})\right|_{\bar{n}_{\rm S}=\sqrt{\delta}c_{\rm cov}/\sqrt{n}}}{\sqrt{n}\log n}&=c_{\rm cov}\sqrt{\delta}c_{\rm rel,EA}.
\end{align}

We derive the expression \eqref{eq:VEA} for $V\left(\hat{\rho}^{BR}\middle\|\hat{\rho}^{B}\otimes\hat{\rho}^{R}\right)$ using the symplectic matrix formalism in Appendix \ref{AppQREV:EA}.
Expansion of $V\left(\hat{\rho}^{BR}\middle\|\hat{\rho}^{B}\otimes\hat{\rho}^{R}\right)$ around $\bar{n}_{\rm S}=0$ in \eqref{eq:AppVEAsmallNS} yields the dominant term $\bar{n}_{\rm S}\log^2\bar{n}_{\rm S}$.
The constant $K_{\rm EA}$ follows from 
\begin{align}
\label{eq:K_EA_limit}\lim_{n\to\infty}\frac{\left.nV\left(\hat{\rho}^{BR}\middle\|\hat{\rho}^{B}\otimes\hat{\rho}^{R}\right)\right|_{{\bar{n}_{\rm S}=\sqrt{\delta}c_{\rm cov}/\sqrt{n}}}}{\sqrt{n}\log^2 n}&=c_{\rm cov}\sqrt{\delta}c_{\mathrm{rel},\text{EA}}.
\end{align}
In Appendix \ref{AppQREV:QEA} we show that $Q\left(\hat{\rho}^{BR}\middle\|\hat{\rho}^{B}\otimes\hat{\rho}^{R}\right)=\mathcal{O}\left(\bar{n}_{\rm S}\log^4\bar{n}_{\rm S}\right)$.
We complete the proof by observing that $\frac{\left[Q\left(\hat{\rho}^{BR}\middle\|\hat{\rho}^{B}\otimes\hat{\rho}^{R}\right)\right]^{3/4}}{V\left(\hat{\rho}^{BR}\middle\|\hat{\rho}^{B}\otimes\hat{\rho}^{R}\right)}=\mathcal{O}\left(n^{1/8}\log n\right)$ dominates the remainder $C_n$ in Lemma \ref{lemma:Mbound}.
\end{IEEEproof}

We are now ready to prove Theorem \ref{th:ea}.

\begin{IEEEproof}[Proof (Theorem \ref{th:ea})]\textbf{Achievability:} 
Dividing both sides of \eqref{eq:M_EAlb} by $\sqrt{n\delta}\log n$ and taking the limit yields the achievable lower bound.

\textbf{Converse:}
Let Alice and Bob have access to respective systems $S$ and $R$ of shared  infinite-dimensional bipartite entangled resource state $\hat{\rho}^{S^mR^m}$, with $m$ arbitrary.
Consider a sequence of codes such that the decoding error probability $\epsilon_n\to0$ as $n\to\infty$.
Then:
\begin{align}
\label{eq:fano_EA}M_{\text{EA}}(1-\epsilon_n)-1&\leq I\left(X^{(n)};\check{X}^{(n)}\right)\\
\label{eq:ea_bound}&\leq n C_{\text{EA}}(\bar{n}_{\rm S};\eta,\bar{n}_{\rm B}),
\end{align}
where $I\left(X^{(n)};\check{X}^{(n)}\right)$ is the mutual information between random variables $X^{(n)}$ and $\check{X}^{(n)}$ corresponding to Alice's message and Bob's decoding of it, \eqref{eq:fano_EA} follows from  Fano's inequality \cite[Th.~2.10.1]{cover02IT}, \eqref{eq:ea_bound} is the entanglement-assisted capacity bound \cite{giovannetti03broadband}.
Substitution of \eqref{eq:ea_cap} and \eqref{eq:ns} into \eqref{eq:ea_bound}, division of both sides by $\sqrt{n\delta}\log n$, and the limit in \eqref{eq:L_EA_limit} yields the converse. 
\end{IEEEproof}

\section{Discussion and conclusion}
\label{sec:discussion}

We derived the quantum-secure covert capacity for the bosonic channel with and without entanglement assistance, closing an important gap from \cite{bullock20discretemod}.
Since entanglement assistance particularly benefits the low-SNR regime, it is expected to improve covert capacity.
Surprisingly, it alters the fundamental scaling law for covert communication from $\mathcal{O}(\sqrt{n})$ to $\mathcal{O}(\sqrt{n}\log n)$ covert bits reliably transmissible in $n$ channel uses.
Next, we outline follow-on questions.

\subsection{Amount of shared resource}
\label{sec:amount_resources}
The resource state $\left(\hat{\rho}^{BR}\right)^{\otimes m}$ employed in the proofs of Lemmas \ref{lemma:noEAfiniteblocklengthn} and \ref{lemma:EAfiniteblocklengthn} is quite large: $m=n2^M$.
This is especially onerous for the entanglement-assisted communications due to the massive costs associated with generating and storing such large entangled states.
While the proofs in \cite{oskouei18unionbound} rely on the state of size $m=n2^M$, we note that our structured receiver design for entanglement-assisted communication \cite{guha20eajdr-isit} (discussed next) uses $m=n$ TMSV states.
The quantum channel resolvability approach\cite{hayashi2003general,nagaoka2007information} should be investigated for reducing $m$ to $\mathcal{O}(\sqrt{n})$, as was done in \cite{bloch15covert} for classical covert communication.

\subsection{Structured receiver for entanglement-assisted covert communication}
\label{sec:jdr}
The sequential decoding strategy from \cite{wilde17ppc,oskouei18unionbound} used by Bob in the proof of Lemma \ref{lemma:EAfiniteblocklengthn} does not correspond to any known receiver architecture. 
In fact, despite the entanglement-based enhancement of classical communication capacity being known for over two decades \cite{bsst02eacap,giovannetti03broadband}, a strategy to achieve the full gain has been elusive until our recent work on the structured receiver for entanglement-assisted communication in \cite{guha20eajdr-isit}.
The receiver in \cite{guha20eajdr-isit} combines insights from the sum-frequency generation receiver proposed for a quantum illumination radar \cite{zhuang17ff-sfg,Shi2019-qk} and the Green Machine receiver proposed for attaining superadditive communication capacity over the bosonic channel \cite{Guha2011-bn}. The resulting structured receiver design realizes the logarithmic scaling gain from entanglement assistance at low SNR.
Consider the approximation \cite[App.~A.2]{guha20eajdr-isit} of this receiver's achievable rate (in bits/mode):
\begin{align}
\label{eq:REMapprox}R_{\rm sr}&\approx \frac{\eta \bar{n}_{\rm S}\gamma}{2(1+(1-\eta)\bar{n}_{\rm B})}\left[\log\left[\frac{w}{\bar{n}_{\rm S}}\right]-\log\left[\ln\left[\frac{w}{\bar{n}_{\rm S}}e\right]\right]-g\left[\frac{2(1-\eta)\bar{n}_{\rm B}(1+(1-\eta)\bar{n}_{\rm B})}{v\eta\gamma}\right]\right],
\end{align}
where $\gamma=1-e^{-2\left(1+(1-\eta)\bar{n}_{\rm B}\right)}$, $w=\frac{4(1+(1-\eta)\bar{n}_{\rm B})}{v\eta\gamma+4(1-\eta)\bar{n}_{\rm B}(1+(1-\eta)\bar{n}_{\rm B})}$, $g(x)$ is defined in \eqref{eq:g}, and $v\geq1$ is a receiver design parameter.
Fixing Bob's receiver makes the Alice-to-Bob channel a classical DMC, allowing us to follow the achievability approach in \cite{wang15covert} almost exactly and obtain the following approximation to its entanglement-assisted covert capacity:
\begin{align}
\label{eq:Ljdr}L_{\text{EA},\rm sr}&\approx\frac{\eta \gamma c_{\rm cov}}{4(1+(1-\eta)\bar{n}_{\rm B})}\approx\frac{L_{\rm EA}}{2},
\end{align}
where the second approximation is valid when $\bar{n}_{\rm B}\gg 1$.
Evolving the receiver \cite{guha20eajdr-isit} to achieve $L_{\rm EA}$ is an ongoing work.

\subsection{Connection to the scaling law for a special case of covert communication without entanglement assistance}
\label{sec:scaling_connection}
Finally, we describe a curious resemblance of the scaling law for entanglement-assisted covert communication presented here to that for a corner case of classical \cite[Th.~7]{bloch15covert} and classical-quantum \cite[Sec.~VII]{azadeh16quantumcovert-isitarxiv} covert communication without entanglement assistance.
Consider a simplified scenario where Alice has two fixed input states $\hat{\rho}^A_0$ and $\hat{\rho}^A_1$, and $\hat{\rho}^A_0$ is the ``innocent'' state that is not suspicious to Willie (e.g., vacuum). 
Let $\hat{\rho}^B_k=\mathcal{N}_{A\to B}\left(\hat{\rho}^A_k\right)$, $k=\{0,1\}$, where $\mathcal{N}_{A\to B}$ is the Alice-to-Bob channel.
Denote the support of $\hat{\rho}$ by $\supp(\hat{\rho})$, and suppose that $\supp\left(\hat{\rho}^B_1\right)\nsubseteq\supp\left(\hat{\rho}^B_0\right)$.
This allows a measurement which perfectly identifies to Bob the transmission of $\hat{\rho}^A_0$.
If Alice is restricted by the SRL to sending $\hat{\rho}^A_1$ with probability $p_1=\mathcal{O}(1/\sqrt{n})$, $\mathcal{O}(\sqrt{n}\log n)$ covert bits can be reliably transmitted in $n$ channel uses \cite[Sec.~VII]{azadeh16quantumcovert-isitarxiv}.
This scaling law was observed prior to \cite{azadeh16quantumcovert-isitarxiv} in the classical covert DMCs with the analogous properties of the supports for corresponding Bob's output probability distributions \cite[Th.~7]{bloch15covert}.
Exploring this connection could lead to new insights in entanglement-assisted communications.

\section*{Acknowledgement}
The authors are grateful to Mark Wilde for pointing out references \cite{oskouei18unionbound,wilde17ppc} and answering our questions.
The authors also benefited from discussions with Saikat Guha and Quntao Zhuang, as well as comments from Evan Anderson and the anonymous reviewer.
\appendices
\def\thesubsection{\mbox{\thesection-\arabic{subsection}}}     
\def\thesubsectiondis{\arabic{subsection}.}         
\newcommand{\sumvar}{k}
\newcommand{\ro}{\hat{\rho}_0} 
\renewcommand{\ro}{\hat{\rho}_{\bar{n}_{\rm T}}}
\newcommand{\rz}{\hat{\rho}_1} 
\renewcommand{\rz}{\hat{\bar{\rho}}^B}
\newcommand{\ra}{\hat{\rho}_{00}} 
\renewcommand{\ra}{\hat{\rho}^B_{\rm T}(u)}
\newcommand{\rb}{\hat{\rho}_{01}} 
\renewcommand{\rb}{\hat{\rho}^B_{\rm T}(ju)}
\newcommand{\rc}{\hat{\rho}_{10}} 
\renewcommand{\rc}{\hat{\rho}^B_{\rm T}(-u)}
\newcommand{\rd}{\hat{\rho}_{11}} 
\renewcommand{\rd}{\hat{\rho}^B_{\rm T}(-ju)}
\newcommand{\nt}{\bar{n}_{\rm T}} 
\renewcommand{\d}{\mathrm{d}}
\newcommand{\dra}{\nt^{-1}\left((\hat{a} - u)\ra\ + \ra(\hat{a}^\dagger - u)\right)} 
\newcommand{\drba}{-\nt^{-1}\left((j\hat{a} + u)\rb\ - \rb(j\hat{a}^\dagger - u)\right)} 
\newcommand{\drc}{-\nt^{-1}\left((\hat{a} + u)\rc\ + \rc(\hat{a}^\dagger + u)\right)} 
\newcommand{\drd}{\nt^{-1}\left((j\hat{a} - u)\rd\ - \rd(j\hat{a}^\dagger + u)\right)} 
\newcommand{\drz}{\dra \\  \drba \\  \drc \\ + \drd} 
\newcommand{\ah}{\hat{a}} 
\newcommand{\ad}{\hat{a}^\dagger} 
\newcommand{\so}{{\hat{\sigma}_0}^{-1}(s)} 
\newcommand{\im}{j} 
\newcommand{\rw}{\hat{\rho}_{\wedge}} 
\newcommand{\rv}{\hat{\rho}_{\vee}} 
\newcommand{\rmid}{\hat{\rho}_{-}} 
\renewcommand{\a}{\mathrm{A}}
\renewcommand{\b}{\mathrm{B}}
\newcommand{\dfive}{\frac{\d^5\rz}{{\d}u^5}}
\newcommand{\dfour}{\frac{\d^4\rz}{{\d}u^4}}
\newcommand{\dthr}{\frac{\d^3\rz}{{\d}u^3}}
\newcommand{\dtwo}{\frac{\d^2\rz}{{\d}u^2}}
\newcommand{\done}{\frac{\d\rz}{{\d}u}}
\newcommand{\qone}{\hat{q}_1}
\newcommand{\qtwo}{\hat{q}_2}
\newcommand{\qthr}{\hat{q}_3}
\newcommand{\qfour}{\hat{q}_4}
\newcommand{\rx}{\hat{\rho}_x}
\renewcommand{\rx}{\hat{\rho}^B_{\rm T}(y)}
\newcommand{\qvar}{V(\rx||\rz)}
\newcommand{\qent}{D(\rx||\rz)}
\newcommand{\sz}{\hat{\sigma}_1^{-1}(s)}
\newcommand{\sx}{\hat{\sigma}_x^{-1}(s)}
\newcommand{\intsx}{\frac{1}{\ln 2}\int_0^1\d{s}\sx\frac{\d\rx}{\d{u}}\sx}
\newcommand{\intsz}{\frac{1}{\ln 2}\int_0^1\d{s}\sz\frac{\d\rz}{\d{u}}\sz}
\newcommand{\intsxu}{\frac{1}{\ln 2}\int_0^1\d{s}\sx\left.\frac{\d\rx}{\d{u}}\right\rvert_{u=0}\sx}
\newcommand{\intszu}{\int_0^1\d{s}\sz\left.\frac{\d\rz}{\d{u}}\right\rvert_{u=0}\sz}
\newcommand{\intsxt}{\int_0^1s\d{s}\sx\frac{\d\rx}{\d{u}}\sx\frac{\d\rx}{\d{u}}\sx}
\newcommand{\inttsx}{\int_0^1\d{s}\sx\frac{\d^2\rx}{\d{u^2}}\sx}
\newcommand{\inttsz}{\int_0^1\d{s}\sz\frac{\d^2\rz}{\d{u^2}}\sz}
\newcommand{\intszt}{\int_0^1s\d{s}\sz\frac{\d\rz}{\d{u}}\sz\frac{\d\rz}{\d{u}}\sz}
\newcommand{\intsxo}{\int_0^1\d{s}\so\left.\frac{\d\rx}{\d{u}}\right\rvert_{u=0}\so}
\newcommand{\intsoa}{\int_0^1\d{s}\so\ro\ah\so}
\newcommand{\intaso}{\int_0^1\d{s}\so\ah\ro\so}
\newcommand{\intsoad}{\int_0^1\d{s}\so\ro\ad\so}
\newcommand{\intadso}{\int_0^1\d{s}\so\ad\ro\so}
\renewcommand{\ket}[1]{\left| #1 \right>} 
\renewcommand{\bra}[1]{\left< #1 \right|} 
\newcommand{\inV}{\hat{R}}

\section{Taylor series expansion of Holevo information and its variance for QPSK modulation}
\subsection{Preliminaries}
In order to prove Theorem 1, we must characterize the behavior of the Holevo information and its variance as a function of the transmitted mean photon number per mode $\bar{n}_\mathrm{S}$ for QPSK. Since the closed-form expressions for \eqref{eq:chidef} and \eqref{eq:vchidef} are unknown, we use Taylor's theorem:
\begin{lemma}[Taylor's theorem]\label{lemma:taylor}
If $f(x)$ is a function with $k+1$ continuous derivatives on the interval 
  $[v,w]$, then
\begin{align}
\nonumber f(w)=&f(v)+f'(v)(w-v)+\ldots+\frac{f^{(k)}(v)}{k!}(w-v)^{k}+R_{k+1}(w)
\end{align}
where $f^{(k)}(x)$ denotes the $k^{\text{th}}$ derivative of $f(x)$, and the Lagrange form of remainder is $R_{k+1}(w)=\frac{f^{(k+1)}(\xi)}{(k+1)!}(w-v)^{k+1}$ with $\xi$ satisfying $v\leq \xi\leq w$.
\end{lemma}
To evaluate the Taylor series expansion, we use the following lemmas where $\hat{A}(t)$ and $\hat{B}(t)$ are non-singular operators parameterized by $t$, and where $\hat{I}$ is the identity operator. 
\begin{lemma}[{\cite[Th.~6]{haber18matrixexplog}}]
\label{lemma:dlogint}
$\frac{\mathrm{d}}{\mathrm{d}t}\ln \hat{A}(t)=\int_0^1\mathrm{d}s\left[s\hat{A}(t)+(1-s)\hat{I}\right]^{-1}\frac{\mathrm{d}\hat{A}(t)}{\mathrm{d}t}\left[s\hat{A}(t)+(1-s)\hat{I}\right]^{-1}$.
\end{lemma}
\begin{lemma}[{\cite[lemma in Sec.~4]{haber18matrixexplog}}]
\label{lemma:dinv}
$\frac{\mathrm{d}}{\mathrm{d}t}\hat{B}^{-1}(t)=-\hat{B}^{-1}(t)\frac{\mathrm{d}\hat{B}(t)}{\mathrm{d}t}\hat{B}^{-1}(t)$.
\end{lemma}

\subsection{Holevo information for quadrature phase shift keying}
\label{app:chiQPSK}
Here, we derive the Taylor series expansion of the Holevo information defined in \eqref{eq:chidef} for QPSK at the displacement $u=0$. Setting $u=0$ in \eqref{eq:chidef} yields
\begin{align}
\left.\chi\left(\{ p_q(y), \rx\}\right) \right|_{u=0} = \ro\log\ro - \ro\log\ro = 0,
\end{align}
where $\hat{\rho}_{\bar{n}_{\rm T}}$ is the zero mean thermal state defined in \eqref{eq:thermal}, with $\nt=(1-\eta)\bar{n}_{\rm B}$.

Von Neumann entropy is invariant under unitary transformations. Since displacement is a unitary, $S(\rx) = S(\ro)$, implying that $\frac{\d S(\rx)}{\d u}=\frac{\d S(\ro)}{\d u}=0$. 
We now evaluate the derivatives of $S(\rz)$ using Lemma \ref{lemma:dlogint}:
\begin{align}
\label{eq:dsQ}\frac{\d S(\rz)}{\d u} =\mathrm{Tr}\left[-\frac{\mathrm{d}\rz}{\mathrm{d}u}\log\rz-\frac{\rz}{\ln 2}\int_0^1\mathrm{d}s\hat{\sigma}_1^{-1}(s)\frac{\mathrm{d}\rz}{\mathrm{d}u}\hat{\sigma}_1^{-1}(s)\right],
\end{align}
where $\hat{\sigma}_1(s)=s\rz+(1-s)\hat{I}$.
The derivatives of $\ra$, $\rb$, $\rc$, and $\rd$ are as follows
  \cite[Ch.~VI, Eq.~(1.31)]{helstrom76quantumdetect}:
\begin{align}
\label{eq:drho00}\frac{\d\ra}{{\d}u}&=\nt^{-1}\left((\hat{a} - u)\ra\ + \ra(\hat{a}^\dagger - u)\right),\\
\label{eq:drho01}\frac{\d\rb}{{\d}u} &= -\nt^{-1}\left((j\hat{a} + u)\rb\ - \rb(j\hat{a}^\dagger - u)\right),\\
\label{eq:drho10}\frac{\d\rc}{{\d}u} &= -\nt^{-1}\left((\hat{a} + u)\rc\ + \rc(\hat{a}^\dagger + u)\right),\\
\label{eq:drho11}\frac{\d\rd}{{\d}u} &= \nt^{-1}\left((j\hat{a} - u)\rd\ - \rd(j\hat{a}^\dagger + u)\right),
\end{align}
where $\hat{a}^\dagger$ and $\hat{a}$ denote the creation and annihilation operators, respectively. Thus,
\begin{align}
\label{eq:drzQ} \frac{\d \rz}{\d u} = \sum_{y\in \mathcal{Q}}p_q(y)\frac{\d \rx}{\d u} =\frac{1}{4}\left(\frac{\d\ra}{{\d}u} + \frac{\d\rb}{{\d}u} + \frac{\d\rc}{{\d}u} + \frac{\d\rd}{{\d}u}\right).
\end{align}
Setting $u=0$ in \eqref{eq:drzQ} yields
$\left.\frac{\d \rz}{\d u}\right\rvert_{u=0} = 0$.
Since both terms in \eqref{eq:dsQ} are zero when $u=0$, $\left.\frac{\d S(\rz)}{\d u}\right\rvert_{u=0} =0$.
Using Lemma \ref{lemma:dinv}, the second derivative of $S(\rz)$ with respect to $u$ is as follows:
\begin{align}
\frac{\d^2 S(\rz)}{\d u^2} =& \mathrm{Tr}\left[-2\frac{\mathrm{d}\rz}{\mathrm{d}u}\frac{1}{\ln 2}\int_0^1\mathrm{d}s\hat{\sigma}_1^{-1}(s)\frac{\mathrm{d}\rz}{\mathrm{d}u}\hat{\sigma}_1^{-1}(s)\nonumber\right.
\\
&\phantom{\mathrm{Tr}}
+2\frac{\rz}{\ln 2}\int_0^1s\mathrm{d}s\hat{\sigma}_1^{-1}(s)\frac{\mathrm{d}\rz}{\mathrm{d}u}\hat{\sigma}_1^{-1}(s)\frac{\mathrm{d}\rz}{\mathrm{d}u}\nonumber\hat{\sigma}_1^{-1}(s)\\
\label{eq:d2sQ}&\left.\phantom{\mathrm{Tr}}-\frac{\rz}{\ln 2}\int_0^1\mathrm{d}s\hat{\sigma}_1^{-1}(s)\frac{\mathrm{d}^2\rz}{\mathrm{d}u^2}\hat{\sigma}_1^{-1}(s)-\frac{\mathrm{d}^2\rz}{\mathrm{d}u^2}\log\rz\right].
\end{align}
Setting $u=0$ in \eqref{eq:d2sQ} and removing terms containing $\left.\frac{\d \rz}{\d u}\right\rvert_{u=0}$ yields
\begin{align}
\label{eq:d2su0Q}\left.\frac{\d^2 S(\rz)}{\d u^2}\right\rvert_{u=0}=\mathrm{Tr}\left[-\frac{\ro}{\ln 2}\int_0^1\mathrm{d}s\hat{\sigma}_0^{-1}(s)\left.\frac{\mathrm{d}^2\rz}{\mathrm{d}u^2}\right|_{u=0}\hat{\sigma}_0^{-1}(s)-\left.\frac{\mathrm{d}^2\rz}{\mathrm{d}u^2}\right\rvert_{u=0}\log\rz\right],
\end{align}
where  $\hat{\sigma}_0(s)=s\ro+(1-s)\hat{I}$.
Setting $u=0$ in $\frac{\mathrm{d}^2\rz}{\mathrm{d}u^2}$ yields
\begin{align}
\label{eq:d2rzQ}\left.\frac{\mathrm{d}^2\rz}{\mathrm{d}u^2}\right\rvert_{u=0}  = \frac{2}{\nt^2}\left(\ah\ro\ad\right)-\frac{2}{\nt}\left(\ro\right).
\end{align}
Substitution of \eqref{eq:d2rzQ} into \eqref{eq:d2su0Q} yields
\begin{align}
\left.\frac{\d^2 S(\rz)}{\d u^2}\right\rvert_{u=0}=\mathrm{Tr}\left[\vphantom{\left(\frac{2}{\nt^2}\left(\ah\ro\ad\right)-\frac{2}{\nt}\left(\ro\right)\right)}\right.&-\frac{2}{\nt^2}\frac{\ro}{\ln 2}\int_0^1\mathrm{d}s\hat{\sigma}_0^{-1}(s)\hat{a}\ro\hat{a}^\dagger\hat{\sigma}_0^{-1}(s)+ \frac{2}{\nt}\frac{\ro}{\ln 2}\int_0^1\mathrm{d}s\hat{\sigma}_0^{-1}(s)\ro\hat{\sigma}_0^{-1}(s)\nonumber\\
\label{eq:d2su0evalQ}&-\left.\left(\frac{2}{\nt^2}\left(\ah\ro\ad\right)-\frac{2}{\nt}\left(\ro\right)\right)\left(\log\ro\right)\right].
\end{align}
Since $\hat{\sigma}_0(s)$ is diagonal in the Fock state basis,
$\hat{\sigma}_0^{-1}(s)=\sum_{k=0}^\infty(s\tau_k+(1-s))^{-1}\ket{k}\bra{k}$,
where $\tau_k=t_k(\nt)$, defined in \eqref{eq:tn}.
Now,
\begin{align}
\nonumber\int_0^1\mathrm{d}s\hat{\sigma}_0^{-1}(s)\ro\hat{\sigma}_0^{-1}(s)= & \int_0^1\mathrm{d}s\sum_{\sumvar=0}^\infty \tau_\sumvar(s\tau_\sumvar+(1-s))^{-2}\ket{\sumvar}\bra{\sumvar}=\hat{I},
	\\
\int_0^1\mathrm{d}s\hat{\sigma}_0^{-1}(s)\hat{a}\ro\hat{a}^\dagger\hat{\sigma}_0^{-1}(s) = &\int_0^1\mathrm{d}s\sum_{\sumvar=0}^\infty(\sumvar+1)\tau_{\sumvar+1}(s\tau_\sumvar+(1-s))^{-2}\ket{\sumvar}\bra{\sumvar}
	\\=& \frac{\nt}{1+\nt}\sum_{\sumvar=0}^\infty(\sumvar+1)\ket{\sumvar}\bra{\sumvar},
\end{align}
since $\int_0^1\mathrm{d}s(sq+(1-s))^{-2}=\frac{1}{q}$ for $q>0$. 
Thus, the traces of the first two terms in \eqref{eq:d2su0evalQ} cancel and we are left with
\begin{align}
\label{eq:d2evalQ}\left.\frac{\d^2 S(\rz)}{\d u^2}\right\rvert_{u=0}=\mathrm{Tr}\left[-\frac{2}{\nt^2}\left(\ah\ro\ad\right)\log\ro+\frac{2}{\nt}\left(\ro\right)\log\ro\right].
\end{align}
The first term in \eqref{eq:d2evalQ} is written in the Fock state basis as
\begin{align}
-\frac{2}{\nt^2}\left(\ah\ro\ad\right)\log\ro &= -\frac{2}{nt^2}\sum_{k=0}^\infty (k+1)\tau_{k+1}\log\tau_k\ket{k}\bra{k}\\
&=-\frac{2}{\nt^2}\left[\log\nt\sum_{k=0}^\infty k(k+1)\tau_{k+1}\ket{k}\bra{k} \right.
\\
&\phantom{=-\frac{2}{\nt^2}\left[\right.}\left.- \log(1+\nt)\sum_{l=0}^\infty (l+1)^2\tau_{l+1}\ket{l}\bra{l}\right].
\end{align}
Taking the trace and evaluating the sums yields
\begin{align}
\mathrm{Tr}\left[-\frac{2}{\nt^2}\left(\ah\ro\ad\right)\log\ro\right]&=-\frac{2}{\nt^2}\left[2\nt^2\log\nt - 2\nt^2\log(1+\nt) - \nt\log(1+\nt)\right]\\
\label{eq:d2.1evalQ}&=-\frac{2}{\nt}\left[2\nt\log\left(\frac{\nt}{1+\nt}\right) - \log(1+\nt)\right].
\end{align}
The second term in \eqref{eq:d2evalQ} can be written in the Fock state basis as
\begin{align}
\frac{2}{\nt}\left(\ro\right)\log\ro &= \frac{2}{\nt}\left[\sum_{k=0}^\infty \tau_k\log\tau_k \ket{k}\bra{k}\right]\\
&= \frac{2}{\nt}\left[\log\nt\sum_{k=0}^\infty k\tau_k\ket{k}\bra{k} - \log(1+\nt)\sum_{l=0}^\infty (l+1)\tau_l\ket{l}\bra{l}\right].
\end{align}
Taking the trace and evaluating the sums yields
\begin{align}
\mathrm{Tr}\left[\frac{2}{\nt}\left(\ro\right)\log\ro \right]&=\frac{2}{\nt}\left[\nt\log\nt - \nt\log(1+\nt) - \log(1+\nt)\right]\\
\label{eq:d2.2evalQ}&= \frac{2}{\nt}\left[\nt\log\left(\frac{\nt}{1+\nt}\right) - \log(1+\nt) \right].
\end{align}
Summing \eqref{eq:d2.1evalQ} and \eqref{eq:d2.2evalQ} yields
$\left.\frac{\d^2 S(\rz)}{\d u^2}\right\rvert_{u=0} = 2\log\left(1+\frac{1}{\nt}\right)$.
Thus, the first non-zero term in the Taylor series expansion of the Holevo information is
\begin{align}
\left.\frac{1}{2!}\frac{\d^2 \chi\left(\{ p_q(y), \rx\}\right)}{\d u^2}\right\rvert_{u=0} = \log\left(1+\frac{1}{\nt}\right).
\end{align}

\subsection{Holevo information variance for quadrature phase shift keying}
\label{app:vQPSK}

Now we derive the Taylor series expansion of Holevo information variance defined in \eqref{eq:vchidef} for QPSK at the displacement $u=0$. The first two derivatives of $\left[D(\rx||\rz)\right]^2$ are: 
\begin{align}
\frac{\d \left(D(\rx||\rz)\right)^2}{\d u} &= 2\frac{\d D(\rx||\rz)}{\d u}D(\rx||\rz),\\
\frac{\d^2 \left(D(\rx||\rz)\right)^2}{\d u^2} &= 2\frac{\d^2 D(\rx||\rz)}{\d u^2}D(\rx||\rz)+2\left(\frac{\d D(\rx||\rz)}{\d u}\right)^2.
\end{align} 
Since $\left.\frac{\d \left(D(\rx||\rz)\right)^2}{\d u}\right\rvert_{u=0}=\left.\frac{\d^2 \left(D(\rx||\rz)\right)^2}{\d u^2}\right\rvert_{u=0}=0$, $\left[D(\rx||\rz)\right]^2$ contributes nothing to the first two terms of the Taylor series. The first two derivatives of $\left[\chi\left(\{ p_q(y), \rx\}\right) \right]^2$ are: 
\begin{align}
\frac{\d \left[\chi\left(\{ p_q(y), \rx\}\right) \right]^2}{\d u} =& 2\frac{\d\chi\left(\{ p_q(y), \rx\}\right) }{\d u}\chi\left(\{ p_q(y), \rx\}\right) ,\\
\nonumber\frac{\d^2 \left[\chi\left(\{ p_q(y), \rx\}\right) \right]^2}{\d u^2} =& 2\frac{\d^2\chi\left(\{ p_q(y), \rx\}\right) }{\d u^2}\chi\left(\{ p_q(y), \rx\}\right) \\
&+2\left(\frac{\d \chi\left(\{ p_q(y), \rx\}\right) }{\d u}\right)^2.
\end{align}
Note that $\left.\frac{\d \left[\chi\left(\{ p_q(y), \rx\}\right) \right]^2}{\d u} \right\rvert_{u=0} = \left.\frac{\d^2 \left[\chi\left(\{ p_q(y), \rx\}\right) \right]^2}{\d u^2}\right\rvert_{u=0} = 0$. Thus, $\left[\chi\left(\{ p_q(y), \rx\}\right) \right]^2$ does not contribute to the first two terms of the Taylor series.
Next, we evaluate the derivatives of $V(\rx||\rz)$. 
Let $\inV=\log{\rx}-\log{\rz}-D(\rx||\rz)$ be the term inside the square in $\qvar$. Note that $\left.\inV\right\rvert_{u=0} = 0$.
The derivative of $\qvar$ with respect to $u$ is:
\begin{align}
\frac{\d\qvar}{\d{u}}=&{\rm{Tr}}\left[\rx\left[\inV\frac{\d\inV}{\d u}+\frac{\d\inV}{\d u}\inV\right]+\frac{\d\rx}{\d u}\inV^2\right].
\end{align}
Setting $u=0$,
$\left.\frac{\d\qvar}{\d{u}}\right\rvert_{u=0}=0$,
since $\left.\inV\right\rvert_{u=0} = 0$. 
The second derivative with respect to $u$ is:
\begin{align}
\frac{\d^2\qvar}{\d{u^2}}=&{\rm{Tr}}\left[\rx\left[2\left(\frac{\d\inV}{\d u}\right)^2+\inV\frac{\d^2\inV}{\d u^2}+\frac{\d^2\inV}{\d u^2}\inV\right]\right.\\
&\phantom{{\rm{Tr}}\left[\right.}\left.+2\frac{\d\rx}{\d u}\left[\inV\frac{\d\inV}{\d u}+\frac{\d\inV}{\d u}\inV\right] + \frac{\d^2\rx}{\d u^2}\inV^2\right].
\end{align}
Setting $u=0$,
\begin{align}
\label{eq:d2Var0}\frac{\d^2\qvar}{\d{u^2}}\rvert_{u=0}=&{\rm{Tr}}\left[2\rx\left(\left.\frac{\d\inV}{\d u}\right\rvert_{u=0}\right)^2\right].
\end{align}
Using Lemma \ref{lemma:dlogint}, we find that the derivative of $\inV$ with respect to $u$ is 
\begin{align}
\frac{\d\inV}{\d u}=& \intsx - \intsz - \frac{\d\qent}{\d u}.
\end{align}
Setting $u=0$, 
\begin{align}
\left.\frac{\d\inV}{\d u}\right\rvert_{u=0}=&\frac{1}{\ln 2}\intsxo,
\end{align}
since $\left.\frac{\d\rz}{\d{u}}\right\rvert_{u=0} = 0$ and $\left.\frac{\d\qent}{\d u}\right\rvert_{u=0} = 0$.
Substituting this term into \eqref{eq:d2Var0} and expanding yields
\begin{align}
&\frac{\d^2\qvar}{\d{u^2}}\left.\vphantom{\frac{\d^2\qvar}{\d{u^2}}}\right\rvert_{u=0}
\\ &\phantom{222222}= \rm{Tr}\left[2\ro\frac{1}{\ln^2 2}\intsxo\intsxo\right].
\end{align}
Summing over $\mathcal{Q}$ yields: 
\begin{align}
\nonumber\sum_{y\in \mathcal{Q}}\left.\frac{\d^2\qvar}{\d{u^2}}\right\rvert_{u=0} 
\label{eq:qvarsum}&=8\frac{\ro}{\nt^2\ln^2 2}\intsoad\intaso\nonumber\\
&\phantom{=}+8\frac{\ro}{\nt^2\ln^2 2}\intaso\intsoad.
\end{align}
Since $\int_0^1\mathrm{d}s(sq+(1-s))^{-1}(sr+(1-s))^{-1}=\frac{\ln\left(\frac{q}{r}\right)}{q-r}$ for $q,r>0$ and $q\neq r$, 
\begin{align}
\intsoad &= \int_0^1\d{s}\sum_{k=0}^\infty \frac{\tau_k \sqrt{k}\ket{k}\bra{k-1}}{(s\tau_k+(1-s))(s\tau_{k-1}+(1-s))}\\
\label{eq:qvarlog1}&=\nt\ln\left(1+\frac{1}{\nt}\right)\sum_{k=0}^\infty \sqrt{k}\ket{k}\bra{k-1},\\
\intaso &= \int_0^1\d{s}\sum_{k=0}^\infty \frac{\tau_k \sqrt{k}\ket{k-1}\bra{k}}{(s\tau_{k-1}+(1-s))(s\tau_{k}+(1-s))}\\
\label{eq:qvarlog2}&=\nt\ln\left(1+\frac{1}{\nt}\right)\sum_{k=0}^\infty \sqrt{k}\ket{k-1}\bra{k}.
\end{align}
Using \eqref{eq:qvarlog1} and \eqref{eq:qvarlog2}, we find the first term of \eqref{eq:qvarsum} as:
\begin{align}
\nonumber8\frac{\ro}{\nt^2\ln^2 2}&\intsoad\intaso \\
&= \frac{8}{\ln^2 2}\ln^2\left(1+\frac{1}{\nt}\right)\sum_{l=0}^\infty \tau_l\ket{l}\bra{l}\sum_{k=0}^\infty \sqrt{k}\ket{k}\bra{k-1}\sum_{m=0}^\infty \sqrt{m}\ket{m-1}\bra{m}\\
&= \frac{8}{\ln^2 2}\ln^2\left(1+\frac{1}{\nt}\right)\sum_{k=0}^\infty k\tau_k\ket{k}\bra{k}\\
&= \frac{8}{\ln^2 2}\nt\ln^2\left(1+\frac{1}{\nt}\right).
\end{align}
Similarly, the second term of \eqref{eq:qvarsum} is:
\begin{align}
\nonumber8\frac{\ro}{\nt^2\ln^2 2}&\intaso\intsoad\\
&= \frac{8}{\ln^2 2}(1+\nt)\ln^2\left(1+\frac{1}{\nt}\right).
\end{align}
Thus, 
\begin{align}
\nonumber\sum_{y\in \mathcal{Q}}\left.\frac{\d^2\qvar}{\d{u^2}}\right\rvert_{u=0} &= 8(1+2\nt)\log^2\left(1+\frac{1}{\nt}\right).
\end{align}
Normalizing by $p_{\rm q}(y)$ yields the first non-zero term in the Taylor series of \eqref{eq:vchidef}:
\begin{align}
\left.\frac{1}{2!}\frac{\d^2 V_\chi(\{ p_q(y),\rx \})}{\d u}\right\rvert_{u=0} = (1+2\nt)\log^2\left(1+\frac{1}{\nt}\right).
\end{align}

\subsection{Fourth central moment of Holevo information for quadrature phase shift keying}
\label{app:qQPSK}

The closed-form expression for the fourth central moment $Q\left(\hat{\rho}^{BR}\middle\|\hat{\rho}^{B}\otimes\hat{\rho}^{R}\right)$, defined in \eqref{eq:qQ}, is unknown for QPSK. 
One can obtain its Taylor series expansion at the displacement $u=0$, using steps very similar to those in Appendices \ref{app:chiQPSK} and \ref{app:vQPSK}.
Inspecting these steps reveals that $Q\left(\hat{\rho}^{BR}\middle\|\hat{\rho}^{B}\otimes\hat{\rho}^{R}\right)=\mathcal{O}\left(\bar{n}_{\rm S}\right)$.
\section{Quantum relative entropy variance for Gaussian states}
\subsection{Preliminaries}
\label{AppQREV:Intro}
Here we employ the symplectic formalism to derive the quantum relative entropy variance for Gaussian and TMSV-based modulation schemes and analyze its asymptotic behavior for small $\bar{n}_{\rm S}$. 
The quantum relative entropy variance $V(\hat{\rho}\|\hat{\sigma})$ between quantum Gaussian states $\hat{\rho}$, $\hat{\sigma}$ with respective first moments $\vec{\mu}_{\hat{\rho}},\ \vec{\mu}_{\hat{\sigma}}$ and covariance matrices (CMs) $\Sigma_{\hat{\rho}},\ \Sigma_{\hat{\sigma}}$ is \cite[Th.~1]{WildeVariance2017}:
\begin{align}
\label{eq:AppVariance}	V(\hat{\rho} \| \hat{\sigma})&=\frac{1}{2} \trace\left[\left(\Delta \Sigma_{\hat{\rho}}\right)^2 \right]+\frac{1}{8} \trace\left[\left(\Delta \Omega\right)^2 \right]+\vec{\delta}^{T} G_{\hat{\sigma}} \Sigma_{\hat{\rho}} G_{\hat{\sigma}} \vec{\delta},
\end{align}
where $\Delta$ is the difference of the Gibbs matrices $\Delta = G_{\hat{\rho}}-G_{\hat{\sigma}}$, $\vec{\delta}=\vec{\mu}_{\hat{\rho}}-\vec{\mu}_{\hat{\sigma}}$, and $
\Omega = \left[\begin{array}{cc}0_{n \times n} & I_{n \times n} \\ -I_{n \times n} & 0_{n \times n}\end{array}\right]
$
is the symplectic matrix in the $qqpp$ representation ($n$ is the number of modes, $I_{n \times n}$ is the $n \times n$ identity matrix, and $0_{n \times n}$ is an $n \times n$ zero matrix). A Gibbs matrix is
$
G_{\hat{\rho}} =-2 \Omega S_{\hat{\rho}}\left[\operatorname{arccoth}\left(2 D_{\hat{\rho}}\right)\right]^{\oplus 2}S_{\hat{\rho}}^T \Omega,
$
where $S_{\hat{\rho}}$ are the symplectic eigenvectors of $\Sigma_{\hat{\rho}}$, $D_{\hat{\rho}}=\\ \text{diag}\left(\lambda_1,\ldots,\lambda_n,\lambda_1,\ldots,\lambda_n\right)$, $\lambda_i$, $i=1,\ldots,n$ are the symplectic eigenvalues, and $\operatorname{arccoth}(x) = \frac{1}{2} \ln \left(\frac{x+1}{x-1}\right),\ x\in(-\infty,-1) \cup (1,+\infty)$. 
A matrix $S$ is symplectic if it is real and satisfies $S \Omega S^T=S^T \Omega S = \Omega$. The symplectic eigenvalues $\lambda_i$, $i=1,\ldots,n$, for a quantum-mechanical system's CM satisfy $\lambda_i\geq 1/2$

We note that a phase shift does not play any role in calculating the quantum relative entropy variance when $\hat{\sigma}$ is  a thermal product state and the rotation is applied in one of the modes of $\hat{\rho}$. Taking into account the correspondence of the phase shift to an orthogonal symplectic matrix $S_\phi$, the property of symplectic matrices $S_\phi^T \Omega S_\phi = \Omega$, and the cyclic permutation property of trace, one can see that \eqref{eq:AppVariance} remains the same if we used $\Sigma_{\hat{\rho}}$ or $\Sigma_{\hat{\rho}}(\phi)=S_\phi \Sigma_{\hat{\rho}} S_\phi^T$. 

. 

\subsection{Quantum relative entropy variance without entanglement assistance}
\label{app:QREVnonEA}
Consider the ensemble of Gaussian single-mode thermal states $\{\hat{\rho}^E(\vec{y})\}$, $\vec{y} \in \mathbb{R}^2$, with CM $\Sigma_{\text{th}}=\left(\bar{n}_{\rm B}+\frac{1}{2}\right) I_{2\times 2}$,
and first moments $\vec{d}=W\vec{y}+\vec{\nu}$,
where $\bar{n}_{\rm B}$ is the mean number of thermal photons, $W$ is a $2 \times 2$ matrix, and $\vec{y}$, $\vec{\nu}$ are $2$-dimensional vectors. The prior classical distribution $p_{\vec{Y}}(\vec{y})$ of this ensemble is Gaussian with the first moments $\vec{\mu}$ and the CM $\Sigma = \bar{n}_{\rm S} I_{2 \times 2}$:
\begin{align}
\label{eq:AppPrior}	p_{\vec{Y}}(\vec{y})=\frac{\exp \left(-\frac{1}{2}(\vec{y}-\vec{\mu})^{T} \Sigma^{-1}(\vec{y}-\vec{\mu})\right)}{2 \pi \sqrt{ \det\Sigma}}.
\end{align}

The expression for the quantum relative entropy variance $V\left(\{p_{\vec{Y}}(\vec{y}),\hat{\rho}^E(\vec{y})\}\right)$ of the ensemble $\{p_{\vec{Y}}(\vec{y}),\hat{\rho}^E(\vec{y})\}$ has been derived in \cite[Def.~1, Prop.~2]{wilde19secondorderqkd},
\begin{align}
V\left(\{p_{\vec{Y}}(\vec{y}),\hat{\rho}^E(\vec{y})\}\right)&=\frac{1}{2}\trace\left[(\Delta \Sigma_{\text{th}})^{2}\right]+\frac{1}{8}\trace\left[(\Delta \omega)^{2}\right] \nonumber\\
\label{eq:AppVarGaussian}&\phantom{=} +\frac{1}{2}\trace\left[W \Sigma W^{T} G_{E} \Sigma_{\text{th}} G_{E}\right]+\frac{1}{2}\trace\left[\left(W \Sigma W^{T} G_{E}\right)^{2}\right]
\end{align}
where,
$\omega = \left[\begin{array}{cc}   0 & 1\\ -1 & 0 \end{array}\right]$, $G = -2 \omega \operatorname{arccoth}(2 \Sigma_{\text{th}})  \omega$, $\Sigma_{E} =\Sigma_{\text{th}}+2 W \Sigma W^{T}$, $G_{E} =-2 \omega \operatorname{arccoth}\left(2 \Sigma_{E} \right) \omega$, $\Delta =G_{E}-G$.
For $W=I_{2 \times 2}$ (and $\vec{\nu}=0$),
i.e., we assume that the first moments of the Gaussian states are equal to the random vector $\vec{y}$: 
\begin{align}
\nonumber	V\left(\{p_{\vec{Y}}(\vec{y}),\hat{\rho}^E(\vec{y})\}\right) &= \bar{n}_{\rm B} (\bar{n}_{\rm B}+1) \ln^2\frac{\bar{n}_{\rm B}+1}{\bar{n}_{\rm B}}
-2 \bar{n}_{\rm B} (\bar{n}_{\rm B}+1) \ln \left(\frac{\bar{n}_{\rm B}+1}{\bar{n}_{\rm B}}\right) \ln \frac{\bar{n}_{\rm B}+\bar{n}_{\rm S}+1}{\bar{n}_{\rm B}+\bar{n}_{\rm S}}\\
\nonumber &\phantom{=}+\bar{n}_{\rm B} (\bar{n}_{\rm B}+1) \ln^2\frac{\bar{n}_{\rm B}+\bar{n}_{\rm S}+1}{\bar{n}_{\rm B}+\bar{n}_{\rm S}}+
\bar{n}_{\rm S} \left(\bar{n}_{\rm B}+\frac{1}{2}\right) \ln^2\frac{\bar{n}_{\rm B}+\bar{n}_{\rm S}+1}{\bar{n}_{\rm B}+\bar{n}_{\rm S}}\\
\label{eq:AppVarGaussian2} &\phantom{=}+\bar{n}_{\rm S}^2 \ln^2\frac{\bar{n}_{\rm B}+\bar{n}_{\rm S}+1}{\bar{n}_{\rm B}+\bar{n}_{\rm S}}.
\end{align}
The Taylor series expansion of \eqref{eq:AppVarGaussian2} at $\bar{n}_{\rm S}=0$ yields:
\begin{align}
	\label{eq:AppVarGaussianApprox}	V\left(\{p_{\vec{Y}}(\vec{y}),\hat{\rho}^E(\vec{y})\}\right)&=  \bar{n}_{\rm S}\left(\bar{n}_{\rm B}+\frac{1}{2}\right)  \ln^2\frac{\bar{n}_{\rm B}+1}{\bar{n}_{\rm B}}+\mathcal{O}(\bar{n}_{\rm S}^2)
\end{align}

\subsection{Quantum relative entropy variance with entanglement assistance}\label{AppQREV:EA}
Here, the upper mode of a TMSV state may be phase-modulated (which we need not consider per the discussion in App.~\ref{AppQREV:Intro}) and sent through a bosonic channel of transitivity $\eta$ and mean thermal photon number $\bar{n}_{\rm B}$. The lower mode does not change. The mean photon number per mode of the TMSV is $\bar{n}_{\rm S}$. The two-mode output state $\hat{\rho}$ is not displaced (i.e., $\vec{\mu}_{\hat{\rho}}=0$ as no displacements are involved in the TMSV nor the evolution of the state) and its CM is:
\begin{align}
\label{eq:AppVrho}	\Sigma_{\hat{\rho}}&=
\left[\begin{array}{cccc}
	 w_{11} & w_{12} & 0 & 0 \\
	w_{12} & w_{22} & 0 & 0 \\
	0 & 0 & w_{11} & -w_{12} \\
	0 & 0 & -w_{12} & w_{22}
\end{array}\right],
\end{align}
where
\begin{align}
\label{eq:Appw11}	w_{11} & = \left(\bar{n}_{\rm B}+\frac{1}{2}\right) (1-\eta)+\left(\bar{n}_{\rm S}+\frac{1}{2}\right) \eta, \\
\label{eq:Appw12}	w_{12} & = \sqrt{\eta \bar{n}_{\rm S} (\bar{n}_{\rm S}+1)},  \\
\label{eq:Appw22}	w_{22} & =\bar{n}_{\rm S}+\frac{1}{2}.
\end{align}
The CM \eqref{eq:AppVrho} is determined from the CM of the TMSV,
\begin{align}
	\Sigma_{\text{TMSV}} &=
\left[\begin{array}{cccc}
	 \bar{n}_{\rm S}+\frac{1}{2} & \sqrt{\bar{n}_{\rm S} (\bar{n}_{\rm S}+1)} & 0 & 0 \\
	\sqrt{\bar{n}_{\rm S} (\bar{n}_{\rm S}+1)} & \bar{n}_{\rm S}+\frac{1}{2} & 0 & 0 \\
	0 & 0 & \bar{n}_{\rm S}+\frac{1}{2} & -\sqrt{\bar{n}_{\rm S} (\bar{n}_{\rm S}+1)} \\
	0 & 0 & -\sqrt{\bar{n}_{\rm S}(\bar{n}_{\rm S}+1)} & \bar{n}_{\rm S}+\frac{1}{2}
	\end{array}\right],
\end{align}
by applying $X \Sigma_{\text{TMSV}} X^T+Y = \Sigma_{\hat{\rho}}$, where the matrices $X,Y$ describe the thermal loss channel which is applied to the upper mode of the TMSV,
\begin{align}
	X &= \text{diag}\left(\sqrt{\eta},1,\sqrt{\eta},1\right),\\
	Y &= \text{diag}\left((1-\eta)\left(\bar{n}_{\rm B}+\frac{1}{2}\right),0,(1-\eta)\left(\bar{n}_{\rm B}+\frac{1}{2}\right),0\right).
\end{align}
We seek the expression for the quantum relative entropy variance $V(\hat{\rho}\|\hat{\sigma})$, where $\hat{\rho}$ is a non-displaced Gaussian state with the CM in \eqref{eq:AppVrho} and $\hat{\sigma}$ is the product of two non-displaced thermal states with $\hat{\rho}$'s CM from \eqref{eq:AppVrho} where all correlations (off-diagonal elements) are set to zero:
\begin{align}
	\label{eq:AppVsigma} \Sigma_{\hat{\sigma}}&=
\left[\begin{array}{cccc}
	w_{11} & 0 & 0 & 0 \\
	0 & w_{22} & 0 & 0 \\
	0 & 0 & w_{11} & 0 \\
	0 & 0 & 0 & w_{22}
\end{array}\right].
\end{align}
To this end we need the symplectic spectrum of $\Sigma_{\hat{\rho}}$ and $\Sigma_{\hat{\sigma}}$, that is, the symplectic matrices $S_{\hat{\rho}},\ S_{\hat{\sigma}}$ and the diagonal matrices $D_{\hat{\rho}},\ D_{\hat{\sigma}}$, such that $\Sigma_{\hat{\rho},\hat{\sigma}}=S_{\hat{\rho},\hat{\sigma}} \left(D_{\hat{\rho},\hat{\sigma}}\oplus D_{\hat{\rho},\hat{\sigma}}\right) S_{\hat{\rho},\hat{\sigma}}^T$. The CM $\Sigma_{\hat{\sigma}}$ is already in the symplectic diagonal form, with $S_{\hat{\sigma}}=I$ and the symplectic eigenvalues $w_{11},w_{22}\geq 1/2$. For $\Sigma_{\hat{\rho}}$, the symplectic eigenvalues are:
\begin{align}
\label{eq:AppSympVal1}	\lambda_1 &= \frac{1}{2} \left(\sqrt{(w_{11}+w_{22})^2-4 w_{12}^2}+(w_{11}-w_{22})\right),\\
\label{eq:AppSympVal2}	\lambda_2 &= \frac{1}{2} \left(\sqrt{(w_{11}+w_{22})^2-4 w_{12}^2}-(w_{11}-w_{22})\right)
\end{align}
and the symplectic eigenvectors (organized into a symplectic matrix),
\begin{align}
\label{eq:AppSympVec}	S_{\hat{\rho}}&=
	\left[\begin{array}{cccc}
	s_+ & s_- & 0 & 0 \\
	-s_- & -s_+ & 0 & 0 \\
	0 & 0 & s_+ & -s_- \\
	0 & 0 & s_- & -s_+ \\
\end{array}\right],
\end{align}
where,
\begin{align}
\label{eq:Appspm}	s_{\pm} &=\frac{1}{2}\left(w\pm\frac{1}{w}\right),\\
\label{eq:Appw}	w&=\frac{\sqrt{w_{11}-2 w_{12}+w_{22}}}{\sqrt[4]{(w_{11}+w_{22})^2-4 w_{12}^2}}.
\end{align}
Using \eqref{eq:AppSympVal1}, \eqref{eq:AppSympVal2}, \eqref{eq:Appspm}, and \eqref{eq:Appw}, one can verify that $D_{\hat{\rho}}=\diag\left(\lambda_1,\lambda_2,\lambda_1,\lambda_2\right)$ is the symplectic diagonal form of $\Sigma_{\hat{\rho}}$:  since $S_{\hat{\rho}} \Omega S_{\hat{\rho}}^T = S_{\hat{\rho}}^T \Omega S_{\hat{\rho}} = \Omega$, $S_{\hat{\rho}}$ is symplectic, and $S_{\hat{\rho}} D_{\hat{\rho}} S_{\hat{\rho}}^T = \Sigma_{\hat{\rho}}$.

We are now ready to apply \eqref{eq:AppVariance} and find the quantum relative entropy variance: 
\begin{align}
\label{eq:VEA}V(\hat{\rho}\|\hat{\sigma})&=\sum_{i=1}^{9} r_i,
\end{align}
where,
\begin{align}
	\label{eq:Appt1} r_1 &= \left(4 w_{11}^2-1\right) \operatorname{arccoth}^2(2 w_{11})\\
	\label{eq:Appt2} r_2 &= \left(4 w_{22}^2-1\right) \operatorname{arccoth}^2(2 w_{22})\\
	\label{eq:Appt3} r_3 &= \left(2 w_{11}^2+2 (w_{11}-w_{22}) \sqrt{(w_{11}+w_{22})^2-4 w_{12}^2}-4 w_{12}^2+2 w_{22}^2-1\right) \operatorname{arccoth}^2(2 \lambda_1) \\
	\label{eq:Appt4} r_4 &= \left(2 w_{11}^2-2 (w_{11}-w_{22}) \sqrt{(w_{11}+w_{22})^2-4 w_{12}^2}-4 w_{12}^2+2 w_{22}^2-1\right) \operatorname{arccoth}^2(2 \lambda_2) \\
	\label{eq:Appt5} r_5 &= 8 w_{12}^2 \operatorname{arccoth}(2 w_{11}) \operatorname{arccoth}(2 w_{22})\\
	\label{eq:Appt6} r_6 &= A(w_{11},w_{12},w_{22}) \operatorname{arccoth}(2 w_{11}) \operatorname{arccoth}(2 \lambda_1)\ \\
	\label{eq:Appt7} r_7 &= \left(A(w_{11},w_{12},w_{22})-2+8 w_{11}^2-8 w_{12}^2\right) \operatorname{arccoth}(2 w_{11}) \operatorname{arccoth}(2 \lambda_2)\\
	\label{eq:Appt8} r_8 &= \left(A(w_{22},w_{12},w_{11})-2+8w_{22}^2-8 w_{12}^2\right) \operatorname{arccoth}(2 w_{22}) \operatorname{arccoth}(2 \lambda_1) \\
	\label{eq:Appt9} r_9 &= A(w_{22},w_{12},w_{11}) \operatorname{arccoth}(2 w_{22}) \operatorname{arccoth}(2 \lambda_2),
\end{align}
with,
\begin{align}
\label{eq:AppA}	A(w_{11},w_{12},w_{22}) &= 1-4(w_{11}^2-w_{12}^2)+\frac{w_{11} + w_{22} - 4 w_{11} (w_{11}^2 - 3 w_{12}^2) - 4 w_{22} (w_{11}^2 + w_{12}^2) }{\sqrt{(w_{11}+w_{22})^2-4 w_{12}^2}}
\end{align}
and $A(w_{22},w_{12},w_{11})$ given by \eqref{eq:AppA} with swapped $w_{11}$ and $w_{22}$. 

Using \eqref{eq:Appw11}-\eqref{eq:Appw22}, and \eqref{eq:Appt1}-\eqref{eq:Appt9} we find that $	\lim_{\bar{n}_{\rm S} \rightarrow 0} V(\hat{\rho}\|\hat{\sigma}) =0$ 
as expected (the quantum relative entropy between two identical states is always zero). By inspection, the leading terms of $ V(\hat{\rho}\|\hat{\sigma})$ scale as $\bar{n}_{\rm S}\ln^2\bar{n}_{\rm S}$.
Expansion of  $V(\hat{\rho}\|\hat{\sigma})$ around $\bar{n}_{\rm S} =0$ yields:
\begin{align}
\nonumber	V(\hat{\rho}\|\hat{\sigma})&=\bar{n}_{\rm S} \ln^2 \bar{n}_{\rm S} +  \frac{\eta}{(1-\eta)\bar{n}_{\rm B} +1}\bar{n}_{\rm S}\ln^2 \frac{(1-\eta )  \bar{n}_{\rm B}+1}{(1-\eta ) \bar{n}_{\rm B}} \\
\nonumber &\phantom{=}-\frac{2 \eta }{(1-\eta)\bar{n}_{\rm B} +1} \left(\ln \bar{n}_{\rm S}\right) \ln\frac{ \left((1-\eta)\bar{n}_{\rm B} +1\right)\bar{n}_{\rm S}}{ (1-\eta)\bar{n}_{\rm B} }   \\
\nonumber &\phantom{=}-\frac{2 (1-\eta ) (\bar{n}_{\rm B}+1) }{(1-\eta ) \bar{n}_{\rm B}+1} \bar{n}_{\rm S} \left(\ln \bar{n}_{\rm S}\right) \ln \frac{(1-\eta ) (\bar{n}_{\rm B}+1) \bar{n}_{\rm S}}{(1-\eta ) \bar{n}_{\rm B}+1}\\
&\phantom{=}+\frac{(1-\eta ) (\bar{n}_{\rm B}+1)}{(1-\eta ) \bar{n}_{\rm B}+1} \bar{n}_{\rm S} \ln^2 \frac{(1-\eta ) (\bar{n}_{\rm B}+1) \bar{n}_{\rm S}}{(1-\eta ) \bar{n}_{\rm B}+1}+\mathcal{O}\left(\left(\bar{n}_{\rm S} \ln \bar{n}_{\rm S}\right)^2\right)\label{eq:AppVEAsmallNS}.
\end{align}

\subsection{Fourth central moment of quantum relative entropy for Gaussian states}
\label{AppQREV:QEA}
Consider zero-mean quantum Gaussian states $\hat{\rho}$, $\hat{\sigma}$ with $\vec{\mu}_{\hat{\rho}}=\vec{\mu}_{\hat{\sigma}}=0$ and CMs $\Sigma_{\hat{\rho}},\ \Sigma_{\hat{\sigma}}$.
The steps in the derivation of the fourth central moment $Q\left(\hat{\rho}\middle\|\hat{\sigma}\right)$ are similar to those leading to \eqref{eq:AppVariance} in the proof of \cite[Th.~1]{WildeVariance2017}.
Inspection of \cite[App.~B]{WildeVariance2017} yields:
\begin{align}
\label{eq:AppQ}	Q(\hat{\rho} \| \hat{\sigma})&=\sum_{k=1}^5c_k\trace\left[\left(\Delta \Sigma_{\hat{\rho}}\right)^{5-k} \left(\Delta \Omega\right)^{k-1} \right],
\end{align}
where obtaining numerical constants $c_k$, $k=1,\ldots,5$ involves applying the commutation relationships of the quadrature operators and Isselsis' theorem \cite{isserlis18gaussmoments}, as is done to derive $V(\hat{\rho} \| \hat{\sigma})$ in \cite[App.~B]{WildeVariance2017}.
We omit this calculation, as we are interested only in how $Q(\hat{\rho} \| \hat{\sigma})$ scales with $\bar{n}_{\rm S}$ when $\Sigma_{\hat{\rho}}$ and $\Sigma_{\hat{\sigma}}$ are defined in \eqref{eq:AppVrho} and \eqref{eq:AppVsigma}, respectively.
The first-order expansion of $Q(\hat{\rho} \| \hat{\sigma})$ around $\bar{n}_{\rm S}$ yields $Q(\hat{\rho} \| \hat{\sigma})=\mathcal{O}\left(\bar{n}_{\rm S}\ln^4 \bar{n}_{\rm S}\right)$.
The logarithm is taken to the fourth power, since it enters through $\Delta$, which is multiplied at most four times in \eqref{eq:AppQ}.

\bibliographystyle{IEEEtran}
\bibliography{../papers}

\end{document}